\documentclass[a4paper]{jpconf}
\usepackage{graphicx}
\usepackage{bm}% bold math
\usepackage{amsmath}
\usepackage{amssymb}
\newcommand{\cla}{$0.701(5)$}
\newcommand{\clb}{$0.898(5)$}
\newcommand{\clc}{$0.911(5)$}
\newcommand{\cld}{$0.940(5)$}
\newcommand{\vla}{$0.826(5)$}
\newcommand{\vlb}{$1.326(5)$}
\newcommand{\vlc}{$1.182(5)$}
\newcommand{\vld}{$1.335(5)$}
\begin{document}
\title{Berezinskii-Kosterlitz-Thouless transition on regular and Villain types of $q$-state clock models} 
%\author{Tasrief Surungan$^{1,3}$, Freddy P. Zen$^{2,3}$, Anthony G. Williams$^4$ }
%\address{$^1$Department of Physics, Hasanuddin University, Makassar 90245, Indonesia}
%\address{$^2$Department of Physics, Bandung Institute of Technology, Bandung 40132, Indonesia}
%\address{$^3$Indonesian Center for Theoretical and Mathematical Physics (ICTMP),
%Bandung Institute of Technology, Bandung 40132, Indonesia}
%\address{$^4$Special Research Center for the Subatomic Structure of Matter (CSSM),
% The University of Adelaide, Adelaide, SA 5005, Australia }
%\ead{tasrief@unhas.ac.id, fpzen@fi.itb.ac.id, anthony.williams@adelaide.edu.au}
\author{Tasrief Surungan$^{1}$, Shunsuke Masuda$^{2}$,Yukihiro Komura$^{2,3}$,
Yutaka Okabe$^2$}
%\affiliation{
\address{$^1$Department of Physics, Hasanuddin University, Makassar, South Sulawesi 90245, Indonesia}
\address{$^2$Department of Physics, Tokyo Metropolitan University, Hachioji, Tokyo 192-0397, Japan}
\address{$^3$RIKEN, Advanced Institute for Computational Science, 7-1-26, Minatojima-minami-machi, Chuo-ku, Kobe, Hyogo, 650-0047, Japan}
\ead{tasrief@unhas.ac.id, okabe@phys.se.tmu.ac.jp}
\date{\today}
\begin{abstract}
We study $q$-state clock models of regular and Villain types with $q=5,6$
using cluster-spin updates and observed double transitions in each model. 
We calculate the correlation ratio and size-dependent correlation length 
as quantities for characterizing the existence of 
Berezinskii-Kosterlitz-Thouless (BKT) phase and its transitions 
by large-scale Monte Carlo simulations. 
We discuss the advantage of correlation ratio in comparison
to other commonly used quantities in probing BKT transition.  
Using finite size scaling of BKT type transition, 
we estimate transition temperatures and corresponding exponents. 
The comparison between the results from both types revealed that
the existing transitions belong to BKT universality.
\end{abstract}

%\pacs{75.40.Mg, 75.10.Hk, 64.60.De, 05.70.Jk}
%\keywords{Clock model, Villain model, Berezinskii-Kosterlitz-Thouless 
%transitions, Monte Carlo}
%\maketitle

\section{Introduction}
The two-dimensional (2D) $q$-state clock models play an important role 
in the study of phase transitions as they provide links 
between several seminal works \cite{onsager,mermin,Berezinskii,kosterlitz,kosterlitz2}.
The underline $Z_q$ symmetry is a generalization of $Z_2$ Ising model, 
the most studied and the first model in statistical physics 
found to exhibit a phase transition \cite{onsager}. 
The XY model, their continuous counterpart, is the native host 
of a unique phase which is the only type of order allowed 
by the well known Mermin-Wagner theorem \cite{mermin} 
for a  2D pure system with continuous symmetry. 
It is a quasi-long-range order (QLRO) called Berezinskii-Kosterlitz-Thouless 
(BKT) phase \cite{Berezinskii,kosterlitz,kosterlitz2}; essentially a topological excitation 
state in the form of vortex-antivortex pairs. 
The BKT phase is ubiquitous for many system, observed for
example in a planar array of coupled Josephson junctions in
transverse magnetic field \cite{Resnick,Marconi}, liquid crystal \cite{Vink}
and the low-dimensional ferroelectrics \cite{Nahas}.
Interestingly, this type of phase is inherited by the clock models, 
depending on the parameter $q$ which evenly divides $2\pi$ into $q$ equal angles.

Based on the Mermin-Wagner theorem above mentioned, a discrete symmetry is 
important for the existence of a true LRO in 2D systems.
It was theoretically predicted \cite{elitzur,jose} and later confirmed
numerically by several early studies \cite{tobo,challa,yamagata} that
the $q$-state clock models experience two transitions for $q > 4$,
i.e., the lower ($T_1$) and the higher temperature ($T_2$) transitions,
where the BKT phase exists between them.
%This is related to the typical characteristic of the KT phase, 
%namely the existence of vortex-antivortex formation. 
The typical characteristics of this phase is related to
the existence of vortex-antivortex formation.
Mathematically, this exotic pattern is associated with the power-law behavior 
of the spin-spin correlation function. 
Since it is not possible to obtain such a formation out of spins with 
$Z_q$ symmetry for $q \leq 4$, the corresponding models cannot host a BKT phase.
%The degree of discreteness is comparatively large and deters very small excitation. 
A relatively small excitation is not allowed as any possible change
of angular directions is always large.
This leads the low-temperature excitation 
to be occupied by a true LRO, rather than BKT phase. For $q > 4$ 
there is an interplay between the discreteness which tends to create
an LRO at low temperature and the inherited symmetry 
trying to preserve BKT phase. The conformity of both causes
the LRO to exist below the BKT phase. Thereby two transition
temperatures, $T_1 < T_2$, are observed; each corresponds to transition
between LRO and QLRO and between QLRO and a disordered phase. 
The LRO is induced by the discrete symmetry of the clock model.

In recent years, the study of phase transition of $q$-state clock models
has regained a lot of interest, in particular to probe whether
the existing double transitions are BKT-type or 
not \cite{lapilli,hwang,baek01,baek02,yuta,crist,baek13}. 
Analytical results from a related model called Villain model \cite{villain} 
revealed that both transitions are BKT type and interconnected 
%via duality relation \cite{savit} for $q > 4$.  
via duality relation \cite{elitzur} for $q > 4$.  
Although such relation does not exist in the regular clock models, 
the renormalization group study \cite{jose} as well as  
standard Monte Carlo studies \cite{tobo,tomita02} pointed out 
two separate transitions belonging to BKT universality.  
This is in contrast to the report by Lapilli {\it et al.} \cite{lapilli} 
who suggested that the transitions are BKT type only for $q\geq 8$.
A study based on Fisher zero approach claimed that the transitions 
for $q=6$ are not BKT type \cite{hwang}, which could possibly 
be an artifact result from small system sizes \cite{baek01}.
While observing double transition for $q=5,6$, 
%Baek and Minnhagen \cite{baek02} concluded that transitions 
Baek {\it et al.} \cite{baek02,baek13} concluded that transitions 
%are KT type for $q=6$, but not for $q=5$ depending on the
%the interaction potential. With these conflicting results, 
are BKT type for $q=6$, but not for $q=5$ due to the
role played by the supposed residual symmetry $Z_5$.
However, more recent study by Kumano {\it et al.} \cite{yuta} and
Chatelain \cite{crist} are in favor of
the previous scenario that the two transitions belonging to BKT type. 
 With these conflicting results, 
the controversy on the type of phase transitions of $q$-state 
clock models is mainly related to $q=5,6$.  
Further investigation using different approaches is required, 
primarily to study $q$-state clock models for $q=5,6$.  

In this paper, we report our study on these particular models, and try 
to address the controversy.
We perform numerical study on regular (cosine) and Villain types of the
models. The Villain type is quite interesting due to
the existence of exact dual transformation by which 
the lower and higher temperature transitions are exchangeable. Due to the
similarity of the underlining symmetry with the clock models, 
studying Villain model can bring a clear insight 
related to the debated issue; in particular if one uses 
the correlation ratio in characterizing the  BKT transition.
We argue that this quantity is more straightforward and exceptionally
suitable in capturing the pattern of vortex-antivortex binding.

The remaining part of the paper is organized
as follows: Section II describes the models and the method. 
The results are discussed in Section III while Section IV is devoted
to the concluding remarks.  
Supplementary subjects are discussed in Appendices.

\section{Models and Simulation Method}

A $q$-state clock model of regular (cosine) type is defined 
by the following Hamiltonian
\begin{equation}\label{Hc}
 H = -J \sum_{<ij>} \vec s_i \cdot \vec s_j
   = - J \sum_{<ij>} \cos(\theta_i - \theta_j) ,
\end{equation}
where spins $\vec s_i$ on lattice sites are planar spins 
restricted to align in $q$ discrete angles ($\theta = 2n\pi/q$ 
with $n = 1,\cdots,q$).  The sum is over nearest-neighbor sites 
of square lattice ($L \times L$), with periodic boundary conditions. 
The interaction is ferromagnetic ($J > 0$) by which
all spins are aligned in the ground state configuration 
with $q$-fold degeneracy; the ground state energy 
is $E_{\rm gs}= -2NJ$ for the square lattice. 

By rewriting Eq.~(\ref{Hc}) as $H = -J\sum_{<ij>} V (\theta_{ij})$ 
with $\theta_{ij}= \theta_i - \theta_j$, 
we can assign a local Boltzmann factor for the clock model 
as $\exp [\kappa V(\theta_{ij})] = \exp [\kappa \cos(\theta_{ij})]$, 
where $\kappa = \beta J$ with $\beta $ as the inverse temperature.
Villain type of clock model is obtained by introducing a set of 
%auxiliary fields \cite{villain,savit}. 
auxiliary fields \cite{elitzur,villain}. 
For the Villain model, the local Boltzmann factor is given by
\begin{equation}\label{Vil}
  \exp[\kappa V_{\rm Vil}(\theta_{ij})] =  \sum_{m=-\infty}^{\infty} 
  \exp[-\frac{\kappa}{2}(\theta_{ij}-2\pi m)^2]. 
\end{equation}
The result of the summation of Gaussian function in Eq.~(\ref{Vil}) 
over $m$ yields a similar form to the local Boltzmann factor 
of the clock model.  The energy origin had better be shifted as 
$\cos(\theta_{ij})-1$ for comparison. 
As we consider the regular (cosine) and Villain types of the models 
for $q=5,6$, there are four models altogether. 

For a reliable calculation, 
we use the canonical Monte Carlo method with  
cluster spin updates proposed by Swendsen and Wang \cite{sw} and 
adopt the embedded scheme of Wolff \cite{wolff} in constructing 
a cluster for clock spins. 
A cluster spin algorithm is crucial for Monte Carlo study
in order to avoid drawback of single spin update which is frequently
suffering from slowing down near the critical point.
This is done by projecting the spins into a randomly selected orientation, 
out of $q$ possibilities, so that the spins 
are divided into two groups (Ising-like spins).  The embedded scheme 
is essential in carrying out cluster algorithm for
such multi-component spins as planar and Heisenberg spins.
After the projection, the original step of the cluster algorithm is 
performed, i.e., by connecting the spin to 
its nearest neighbors of the same group, with suitable probability 
depending on the local Boltzmann factor \cite{kasteleyn}.  
Spins in the cluster are updated at a time. 
One Monte Carlo step (MCS) is defined as updating all the spins. 

We make measurement for the systems of linear size up to $L=512$. 
In the careful discussion on the helicity modulus, which 
will be given in Appendix A, we treat large enough sizes up to $L=2048$. 
For the calculation of large systems, we use the GPU-based 
algorithm for the Swendsen-Wang algorithm 
\cite{komura_GPU}, which is efficient for high-performance computing. 
Measurement is performed after enough equilibration steps 
($2 \times 10^4$ MCS's). Each data point is obtained from the average 
over several parallel runs, with $4 \times 10^5$ MCS's for each run. 
Typically, 5 measurements are done for each system size 
in order to estimate statistical errors. This means, each
data point is obtained from the average of $2 \times 10^6$
samples.

\section{Results}

\subsection{BKT transitions and Correlation ratio}

A BKT phase is essentially a topological excitation in 
the form of vortex-antivortex pairs \cite{kosterlitz,kosterlitz2}.  
In the case of XY model, the unbinding of these pairs at high temperature 
is assigned as BKT transition which separates the BKT phase 
and paramagnetic phase. 
It corresponds to the $T_2$ transition of the clock 
models which have additional lower-temperature ($T_1$) transition. 
The disappearance of vortex-antivortex pairs following the unbinding
scenario at high temperatures is associated with the vanishing of 
helicity modulus $\Upsilon$ \cite{nelson,minnhagen}.
%Essentially, the helicity modulus tries to capture the spin configuration
%in BKT phase by taking into account the orientation of each spin.
%It is done by algebraically adding the contribution from each spin.
We briefly discuss this quantity in Appendix A and present the result 
for examining the existing BKT phase of the 5-state and the 6-state clock model.
%
%Although this quantity is commonly used to probe BKT transition, it is
%not appropriate for characterizing the $T_1$ transition
%as disappearance of vortex-antivortex pairs at low temperature 
%follows different mechanism. 
Although this quantity is commonly used to probe BKT transition, it is
not quite precise in characterizing the $T_1$ transition
as the disappearance of vortex-antivortex pairs at low temperature
follows different mechanism.
In the case of the clock model, the spin-wave excitations are prohibited 
due to the energy gap of discrete models at low temperatures. 
Thus, the vortex-antivortex pairs disappear at low temperatures, 
%which is the origin of the $T_1$ transition. 
associated with the $T_1$ transition; although the origin
of this transition is due the discrete symmetry of the clock models.
%The conflicting main conclusions reported 
%by Refs. \cite{baek02,baek13} and Refs. \cite{yuta,crist} 
%do indicate a flaw  of helicity
%modulus used for the continuous 
%XY model in probing BKT phase of clock models.
%
The conflicting main conclusions reported
by Refs. \cite{baek02,baek13} and Refs. \cite{crist}
suggest a more accurate way of calculating  the helicity
modulus when dealing  BKT transitions of the clock models \cite{yuta}.

A quantity called correlation ratio \cite{tomita02b} is particularly 
applicable to probing BKT transitions both for the upper and 
the lower transitions, 
and has been used in various cases \cite{ts04,kawa03,kawa07}. 
It is a ratio of correlation functions with different distances.
At a critical point or line (BKT phase), 
the correlation function $g(r)$ 
for an infinite system decays as a power of distance $r$,
\begin{equation}
 g(r) = \left\langle \frac{1}{L^D} \sum_i 
 \vec s_i\cdot \vec s_{i+r} \right\rangle \sim r^{-(D-2+\eta)} .
\end{equation}
where $D$ and $\eta$ are respectively the lattice dimension and decay exponent.
The angular bracket denotes the thermal average. 
It incorporates two length ratios for a finite system 
of linear system size $L$ away from the critical point or line, 
\begin{equation}
 g(r) \sim r^{-(D-2+\eta)} h(r/L, L/\xi)
\end{equation}
with a correlation length $\xi$ for the infinite system.  
The ratio of correlation functions of different distances, 
\begin{equation}\label{gr}
\frac{g(r)}{g(r')} = 
 \left(\frac{r}{r'}\right)^{-(D-2+\eta)}\frac{ h(r/L, L/\xi)}
 { h(r'/L, L/\xi)}\,
\end{equation}
will have a single scaling variable for finite size scaling (FSS),
\begin{equation}
\frac{g(r)}{g(r')} = \tilde f(L/\xi)
\label{scaling}
\end{equation}
if we fix two ratios $r/r'$ and $r/L$. 
In principle, one can take any pair of distances, but for the convenience of numerical calculation
we choose $r=L/2$ and $r'= L/4$.

At the critical region, where the correlation length $\xi$ is infinite,
the correlation ratio, Eq.~(\ref{scaling}), 
does not depend on the system size $L$.  
Thus, it is natural to expect a presence of a crossing point 
for the existence of a continuous transition or 
collapsing curves for that of BKT phase in the plot of 
temperature dependence of correlation ratio for different system sizes.

%--------------------------------fig_R_q6------------------------------------
\begin{figure}[t]
\includegraphics[width=0.48\linewidth]{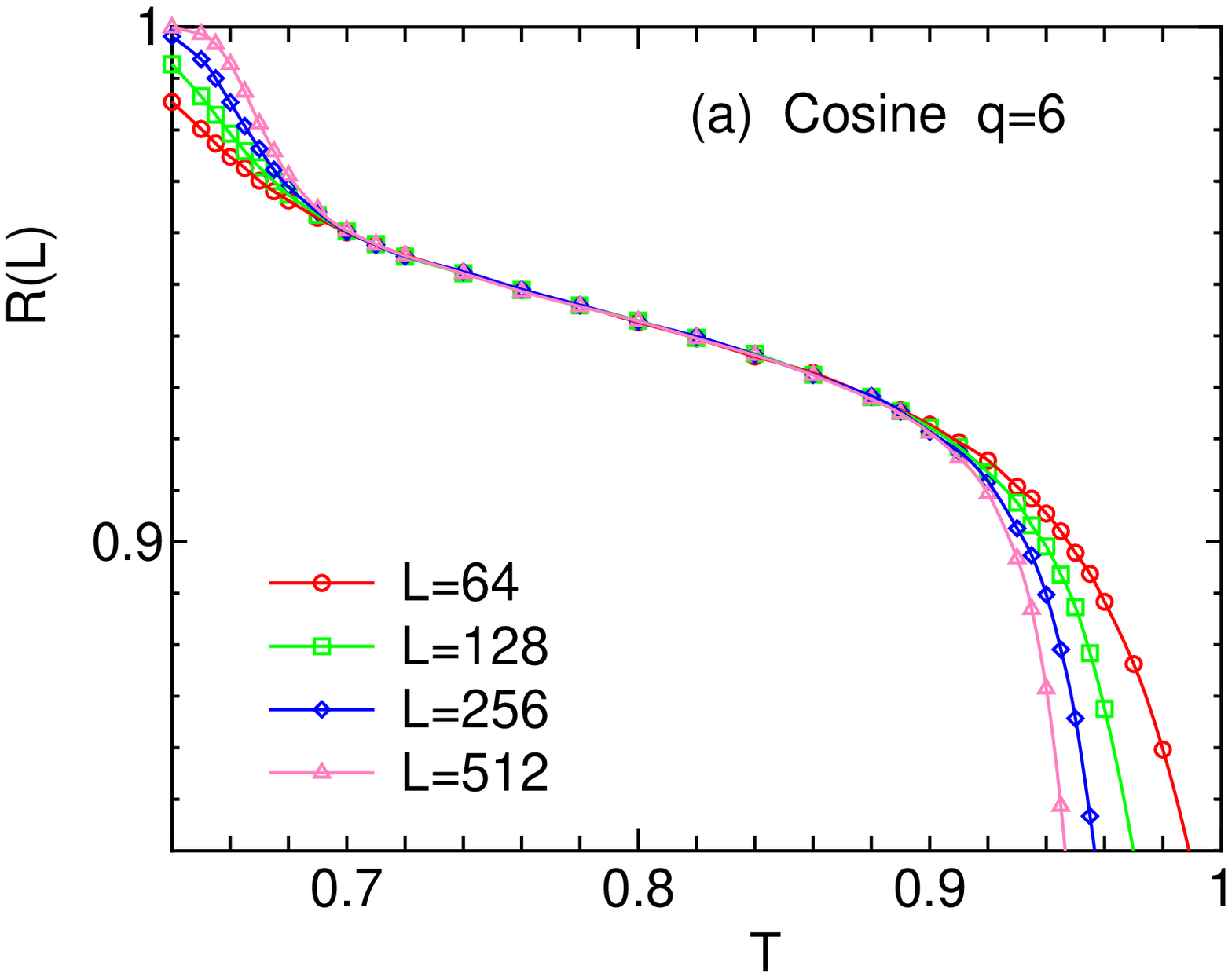}
\includegraphics[width=0.48\linewidth]{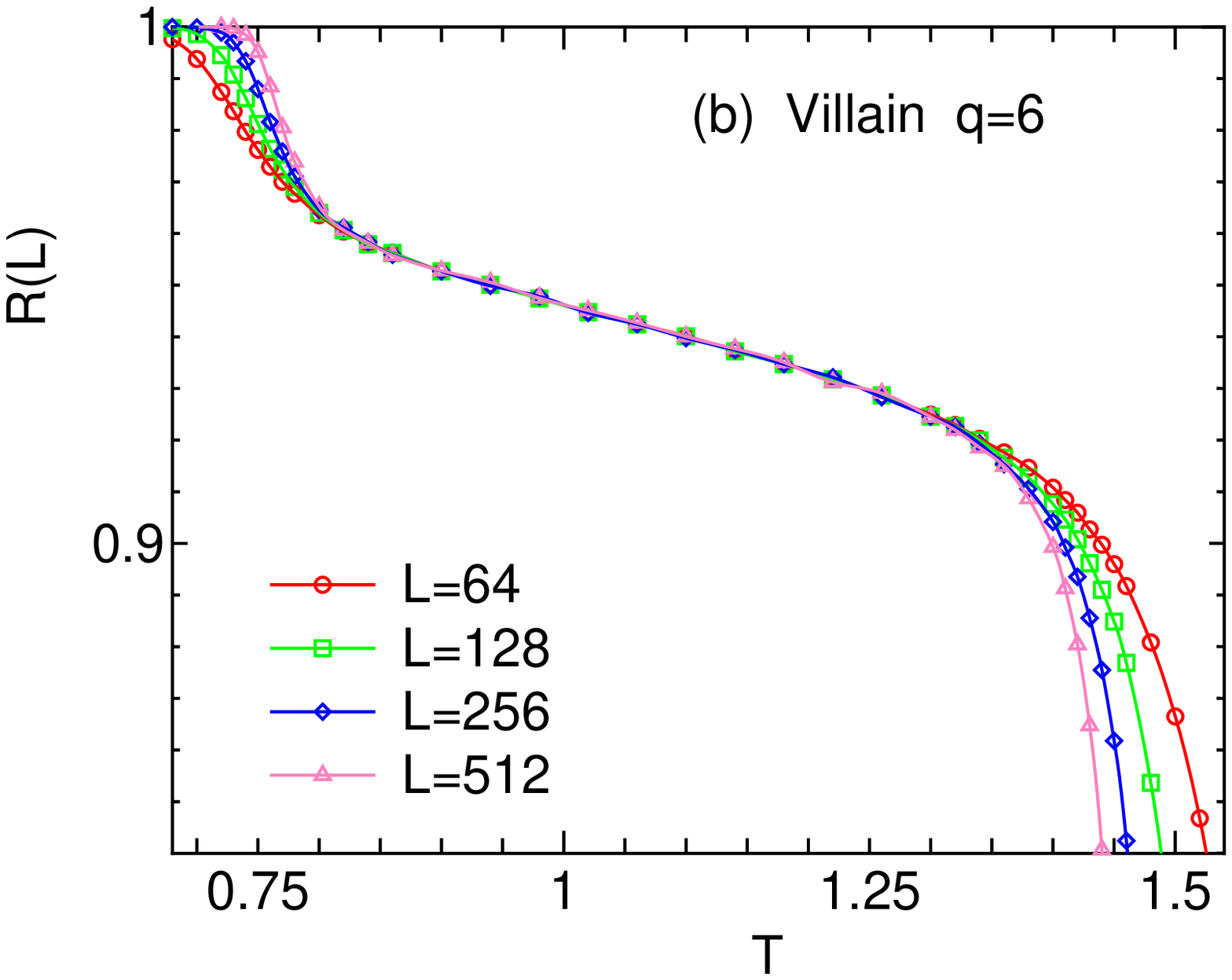}
\caption{(Color online) 
The temperature dependence correlation ratio, 
$R(L)=g(L/2)/g(L/4)$, for (a) regular and (b) Villain types for $q=6$.  
The BKT phase and transitions are respectively indicated by the collapsed lines 
in the intermediate temperature and points where lines start separating.
The error bar, in average, is in the order of symbol size.}
\label{fig_R_q6}
\end{figure}
%------------------------------------------------------------------------------

We plot, as shown in Fig.~\ref{fig_R_q6}, the temperature dependence of 
the correlation ratio, 
\begin{equation}
  R(L) = g(L/2)/g(L/4), 
\end{equation}
for $q=6$ of (a) regular and (b) Villain types, respectively. 
The plots have three typical temperature regimes, i.e., the lower, 
the intermediate and the upper regime. 
Each is associated with LRO, the BKT phase and the paramagnetic phase.  
The BKT phase is indicated by the collapsing curves of different system sizes.
It is a direct consequence of power law behavior of 
correlation function at BKT phase.
The spray out of the curves at lower and higher temperatures signifies
two separated transitions.

%--------------------------------fig_R_q5--------------------------------------
\begin{figure}[t]
\includegraphics[width=0.48\linewidth]{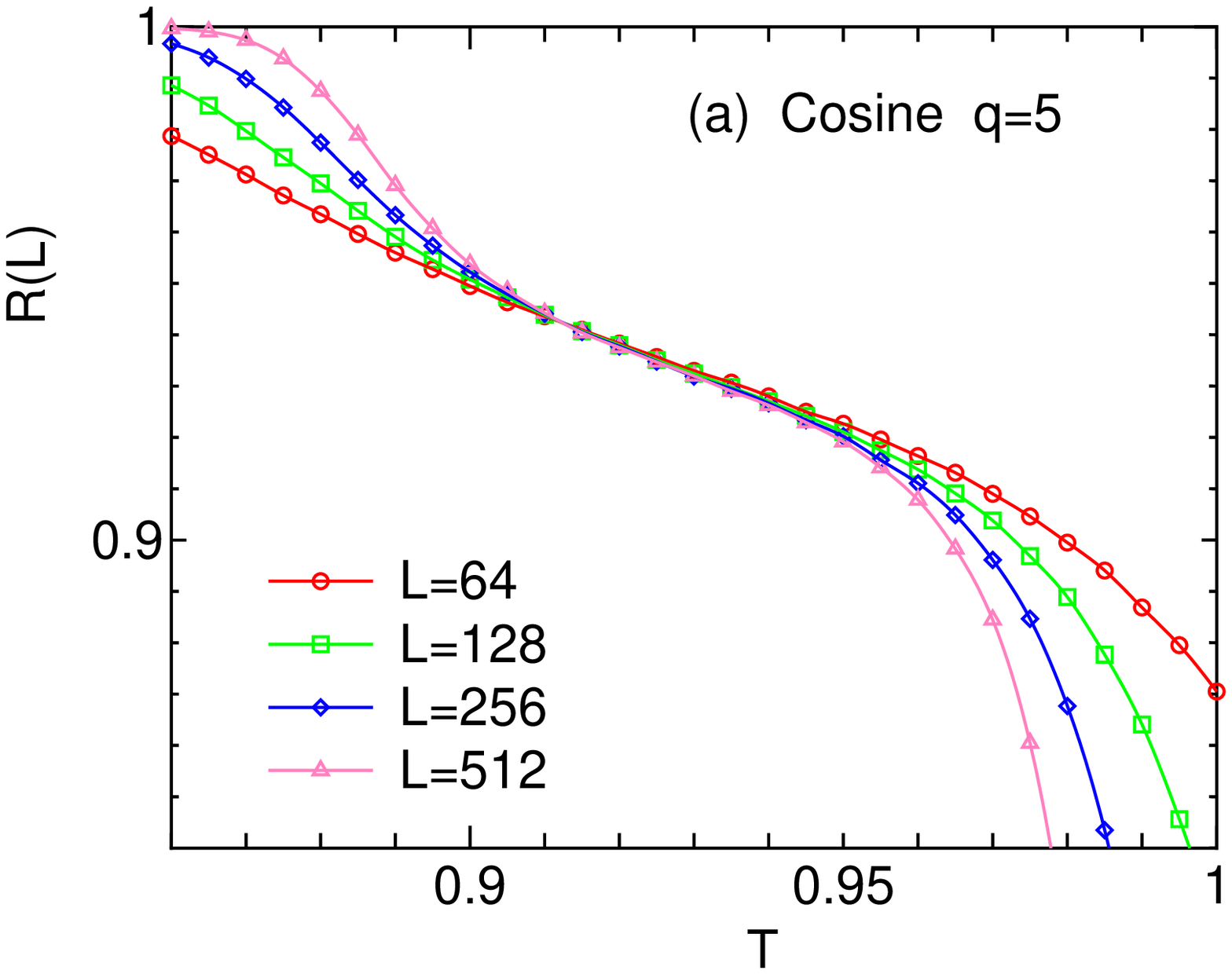}
\includegraphics[width=0.48\linewidth]{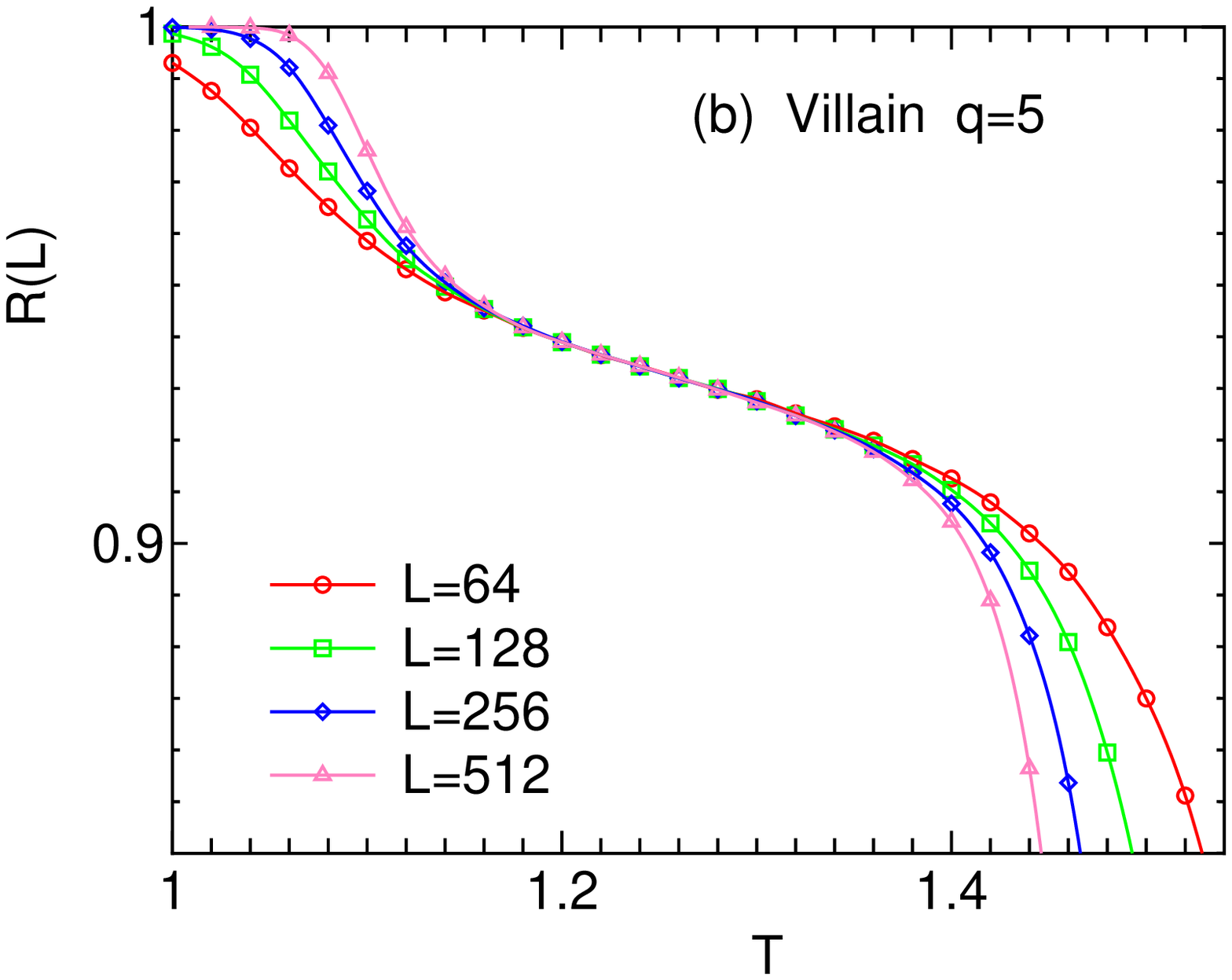}
\caption{(Color online) 
Temperature dependence of correlation ratio for different system sizes
of (a) regular and (b) Villain types for $q=5$. 
The typical pattern for the existence of BKT phase preserves 
except that the temperature range at which curves are collapsing
becomes narrow.}  
\label{fig_R_q5}
\end{figure}
%------------------------------------------------------------------------------

The overall behavior of the correlation ratio, depicted in Fig.~\ref{fig_R_q6}, 
is essentially the same 
for the model of regular type and that of Villain type. 
For the Villain model, 
the upper and the lower transitions are inter-connected via an
exact duality relation \cite{elitzur},
\begin{equation}
  T_1 T_2 =\frac{4\pi^2}{q^2},
\label{DL}
\end{equation}
and both transitions are BKT type \cite{tomita02}. 
Although there is not an exact duality relation for the regular 
clock model, quantitative data shown in Fig.~\ref{fig_R_q6} 
strongly suggest the universality of two transitions for 
regular and Villain types of $q=6$ clock model.  

Analogous to Fig.~\ref{fig_R_q6}, we plot the correlation ratio of 
(a) regular and  (b) Villain types for $q=5$ in Fig.~\ref{fig_R_q5}. 
The typical pattern of correlation ratio indicating the existence of BKT phase
preserves, except that the collapsing regime becomes narrow, which
indicates that the two transitions are getting closer.  
The discreteness suppresses the lower temperature excitation, but it
does not totally rule out the BKT phase.
Since the pattern of regular type is similar to that of Villain model, 
we argue that the transitions of 5-state clock model are also BKT type.  
The precise estimates of transition temperatures and 
the corresponding exponents are presented in the next sub-section.

The excellent performance of correlation ratio in characterizing 
the existence of BKT transition is related to the fact 
that it directly incorporates the correlation function, 
whose algebraic decay at BKT phase is a typical characteristic. 
We note that although the Binder ratio \cite{binder}, 
essentially the moment ratio, has the same FSS property 
as in Eq.~(\ref{scaling}), 
the behavior of the collapsing and spraying out is not good 
due to the large corrections to FSS \cite{tomita02b}. 

We also calculate the size-dependent 2nd-moment correlation length 
for finite system, $\xi(L)$, which is defined as\cite{jose}
\begin{equation}
  \xi(L) = \frac{[\chi(0)/\chi(2\pi/L)-1]^{1/2}}{2\sin(\pi/L)} 
\end{equation}
with
\begin{equation}
  \chi(k) = \frac{1}{N}\left|\left\langle \sum_r s(r) e^{ikr}
            \right\rangle \right|^2
\end{equation}
The quantity $\xi(L)/L$ has an FSS property similar to Eq.~(\ref{scaling}), 
and has been frequently used in spin glass problems \cite{katzgraber}.  
In this case again, the FSS of the Binder ratio \cite{binder} 
is not good compared to that of $\xi(L)/L$. 

%--------------------------------Fig_xi_q6---------------------------------
\begin{figure}[t]
\includegraphics[width=0.48\linewidth]{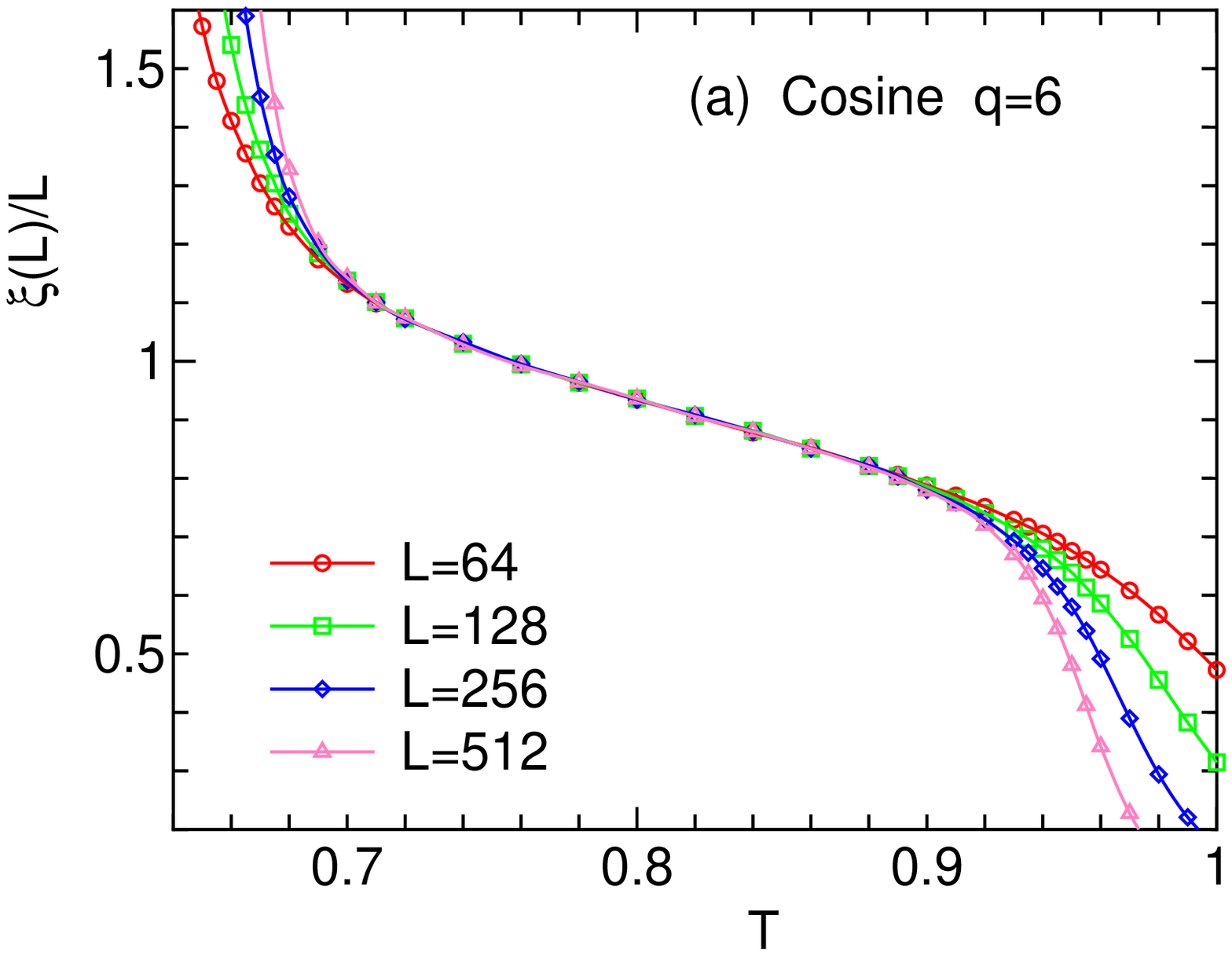}
\includegraphics[width=0.48\linewidth]{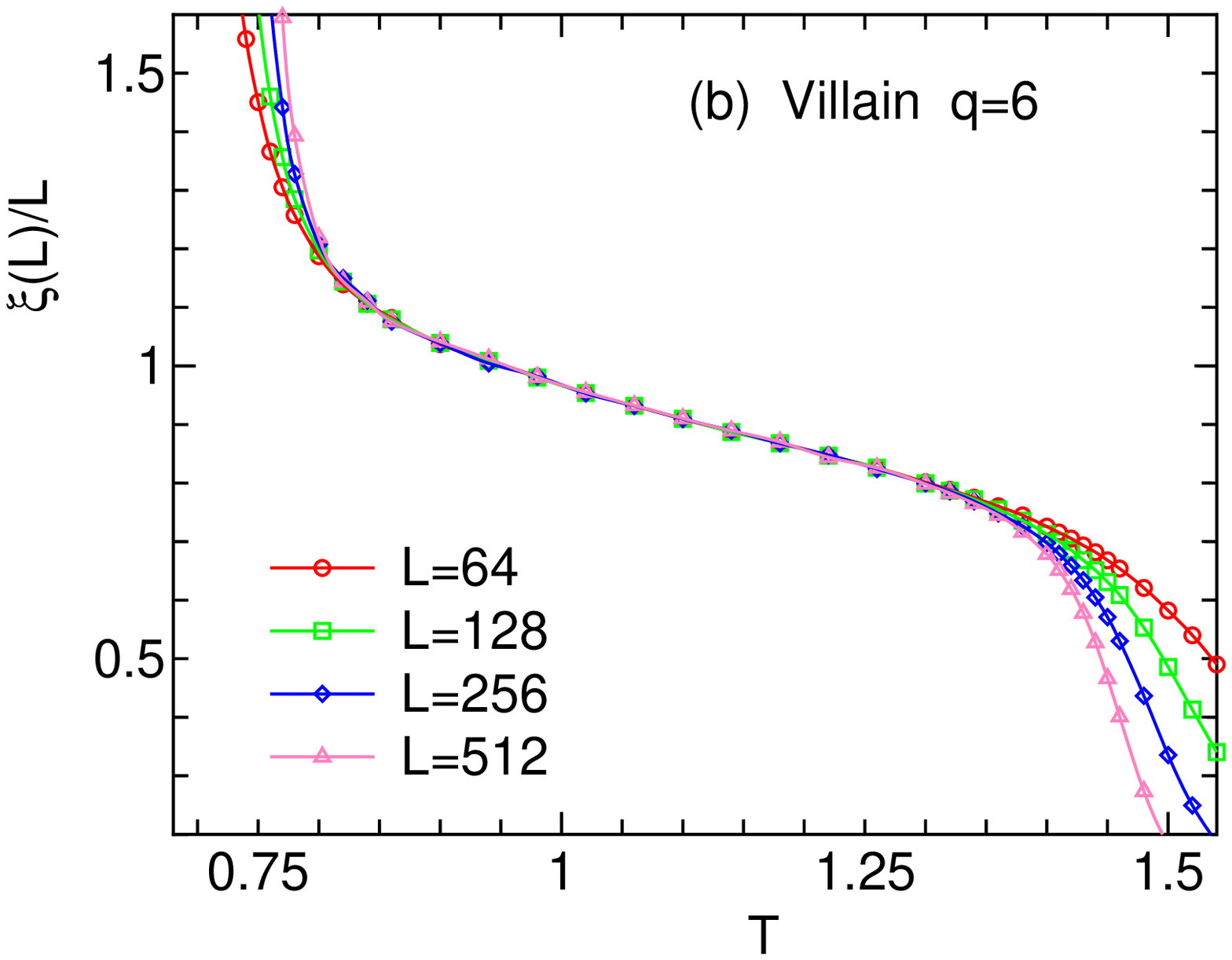}
\caption{(Color online) 
Temperature dependence of $\xi(L)/L$ for different system sizes
of (a) regular and (b) Villain types for $q$=6. 
The typical pattern for the existence
of BKT phase preserves.}
\label{fig_xi_q6}
\end{figure}
%------------------------------------------------------------------------------

\begin{figure}[t]
\begin{center}
\includegraphics[width=0.48\linewidth]{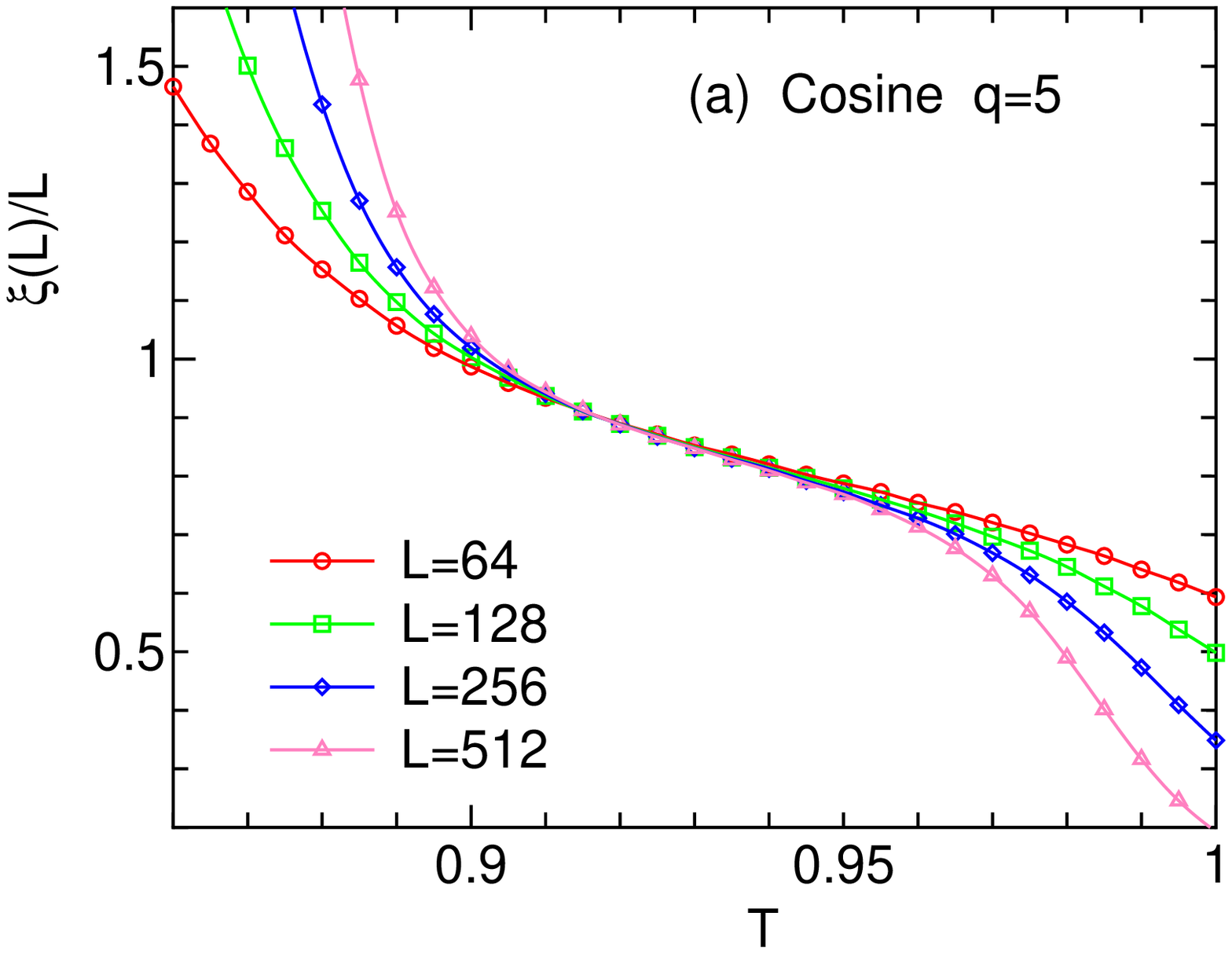}
\includegraphics[width=0.48\linewidth]{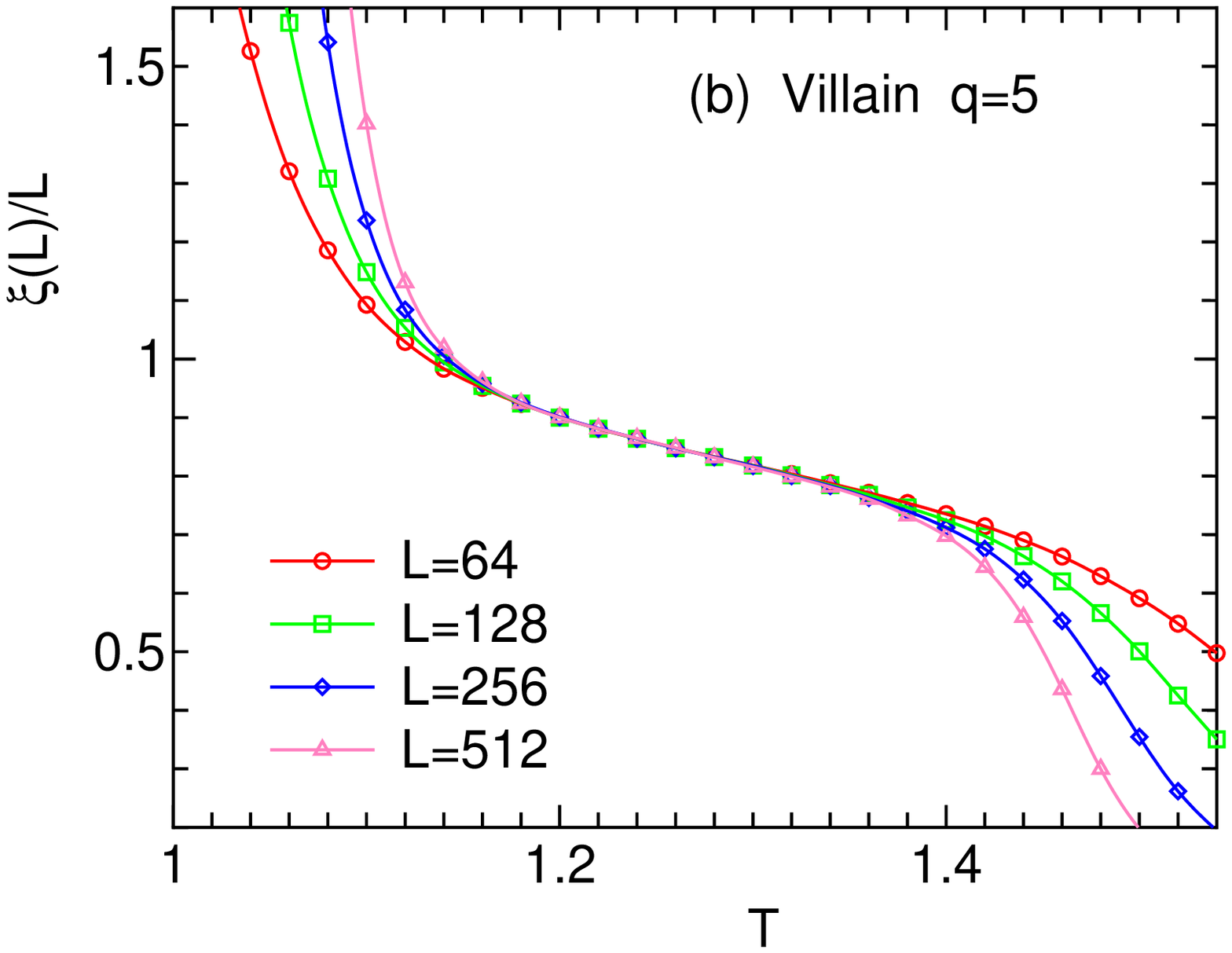}
\caption{(Color online) 
Temperature dependence of $\xi(L)/L$ for different system sizes
of (a) regular and (b) Villain types for $q$=5. 
The typical pattern for the existence
of BKT phase preserves except that the temperature range at which curves are collapsing
becomes narrow.}
\label{fig_xi_q5}
\end{center}
\end{figure}

We plot $\xi(L)/L$ for the regular and Villain types
of $q=6$ in Fig.~\ref{fig_xi_q6}. The collapsing curves at intermediate temperature regime
and the spray out at lower and higher temperatures were observed at both types
of models. This is another indication that the transitions of Villain and
the regular types are of the same universality. 
For $q=5$, shown in Fig.~\ref{fig_xi_q5}, the temperature range in which curves 
are collapsing becomes narrow; it is consistent with what observed 
for the correlation ratio shown in Fig.~\ref{fig_R_q5}. 
All the data and plots presented thus far indicate that 
the transitions of $q$-state clock models of both types for $q=5,6$ 
are universal.

%--------------------------------Fig_xi_q5---------------------------------
%------------------------------------------------------------------------------

\subsection{FSS and the estimate of critical parameters}
FSS is a standard technique in the study of phase transition. 
It is used to extract critical parameters from the finite system sizes.
Here, we present FSS analysis based on diverging behavior 
of correlation length at BKT phase,
\begin{equation}
   \xi \propto \exp (c/\sqrt t)
\end{equation}
with $t = (T - T_{BKT})/T_{BKT}$. We choose $T_{BKT}$ and a constant $c$ 
such that all the data points in Fig. 4 are collapsed on a single curve 
for both the lower transition temperature ($T_1$) and the higher 
transition temperature ($T_2$). 

\begin{figure}[t]
\includegraphics[width=0.494\linewidth]{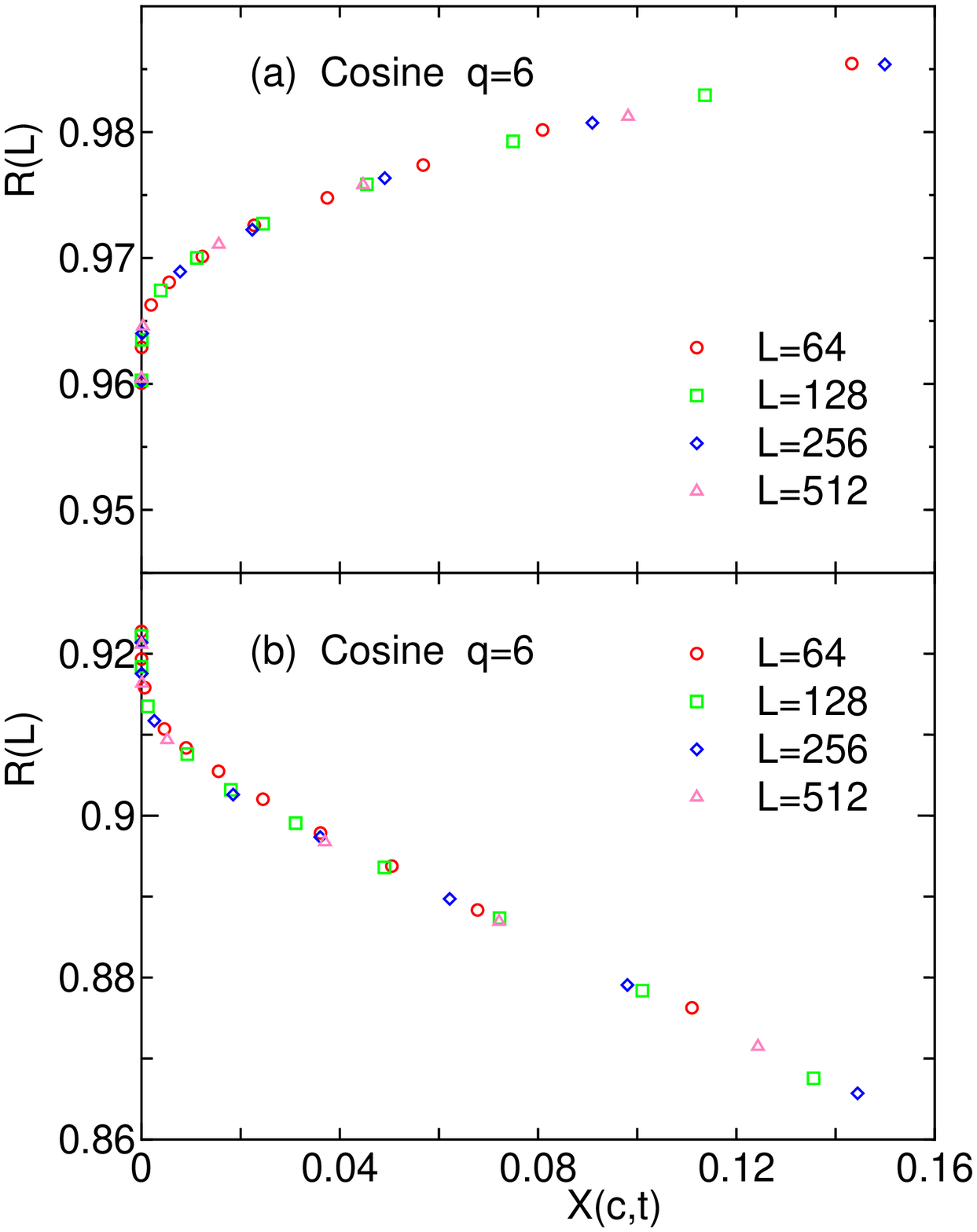}
\includegraphics[width=0.476\linewidth]{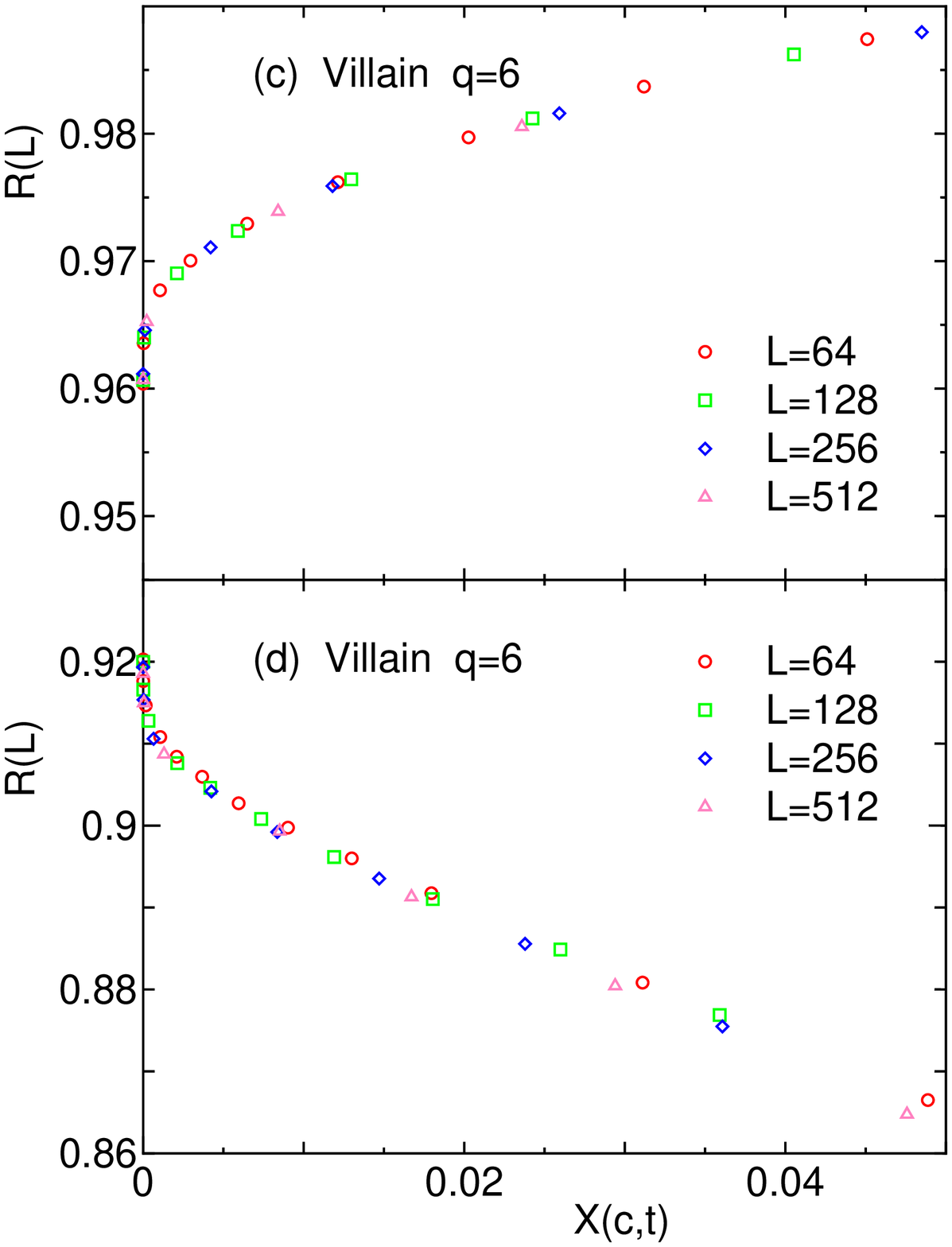}
\caption{(Color online) 
The FSS plot of correlation ratio for the estimate 
of transition temperatures of $q=6$ clock model. 
The scaled variable for the horizontal axis $X(c,t)$ is 
defined by Eq.~(\ref{scaled_variable}).
The estimated values are: (a) $T_1$=\cla\ and (b) $T_2$
=\clb\ for the regular model, and 
(c) $T_1$=\vla\ and (d) $T_2$=\vlb\ for the Villain model.
}
\label{fig_FSS_q6}
\end{figure}

The FSS plot of the correlation ratio, whose original data is given 
in Fig.~\ref{fig_R_q6}, is shown in Fig.~\ref{fig_FSS_q6}, 
associated with the data for $q$=6 of regular and Villain models.  
The plot is obtained by rescaling the horizontal axis, $L/\xi$.
The scaled variable $X(c,t)$ is defined by
\begin{equation}
  X(c,t) = \frac{L}{\exp(c/\sqrt{t})}
         = \frac{L}{\exp(c/\sqrt{|(T-T_{BKT})/T_{BKT}|})};
\label{scaled_variable}
\end{equation}
$c$ and $T_{BKT}$ are $c_1$ and $T_1$ for lower-temperature 
transition, and they are $c_2$ and $T_2$ for higher-temperature 
transition, $c_1$ and $c_2$ being the fitting parameters.
Choosing several pairs of $c_1$ and $T_1$ as well as $c_2$ and $T_2$, 
we obtained the fits for the estimated values. 
Our estimate for $T_1$ and $T_2$ for the regular model are 
respectively \cla\ and \clb; while
for the Villain model $T_1$=\vla\ and $T_2$=\vlb.
The numbers in parentheses are the uncertainty for the last digits. 
The transition temperatures for the Villain model are higher. 
% related to the fact that the configurational energy space 
% is larger due to the presence of auxiliary fields. 

\begin{figure}[t]
\includegraphics[width=0.491\linewidth]{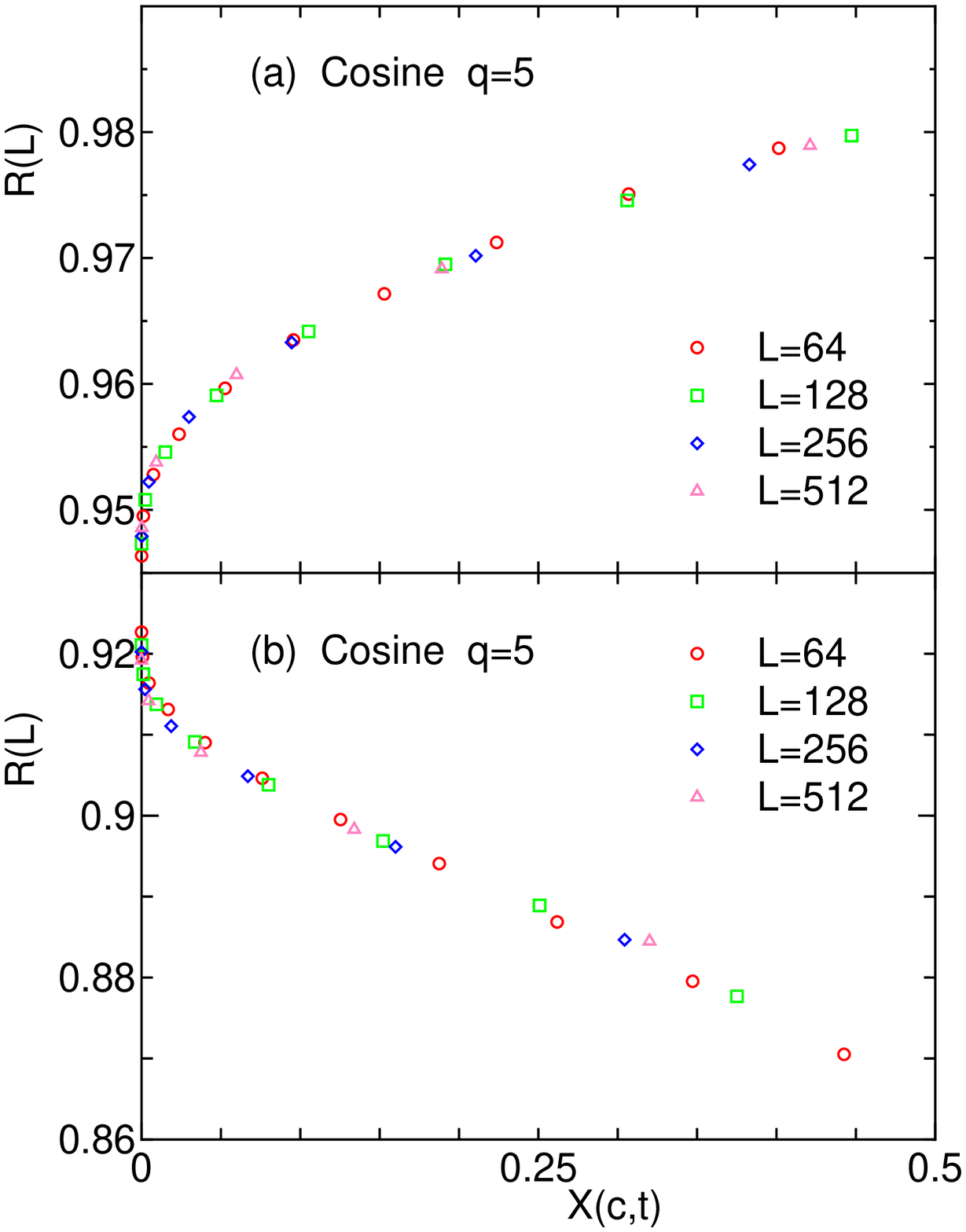}
\includegraphics[width=0.477\linewidth]{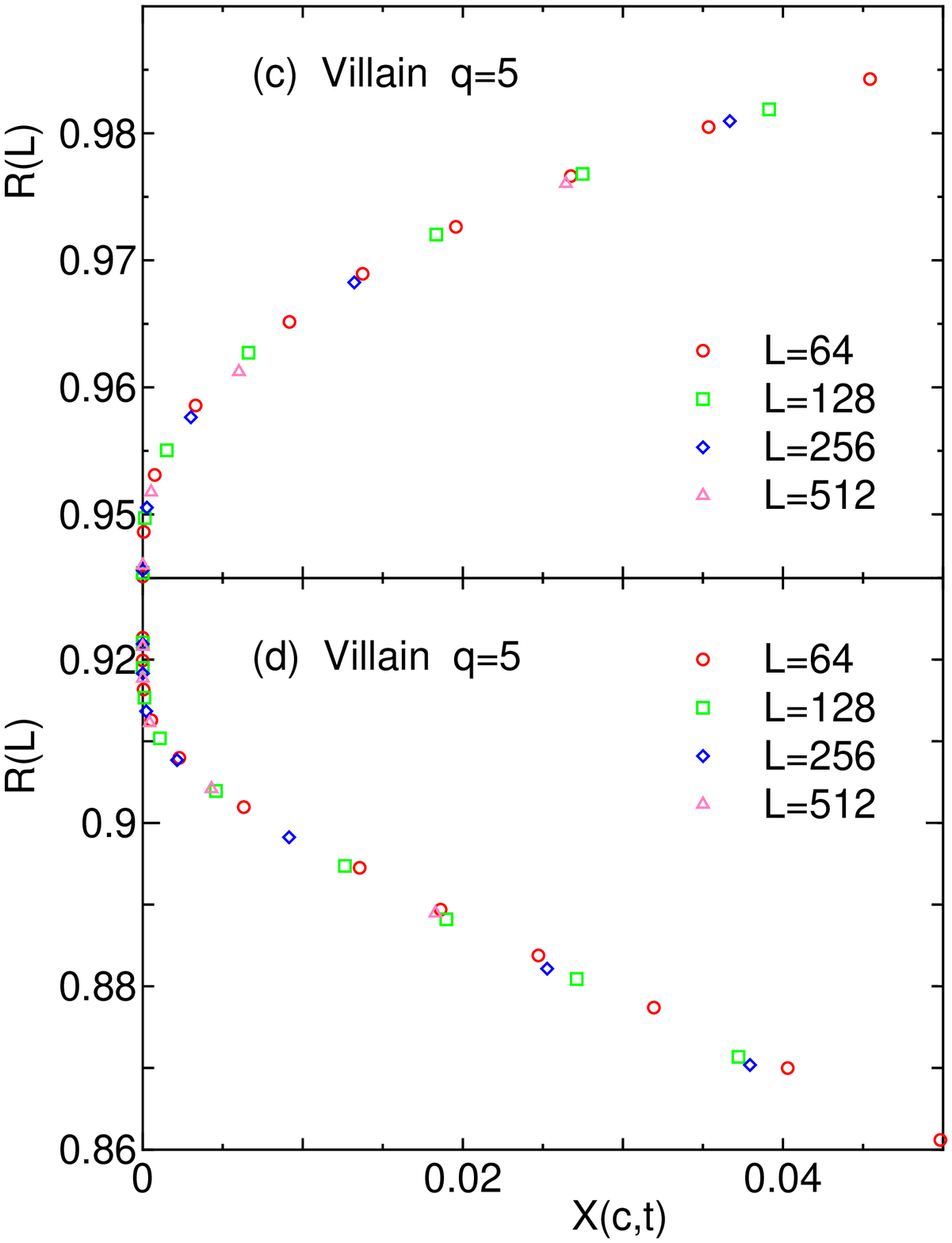}
\caption{(Color online) 
The FSS plot of correlation ratio for the estimate 
of transition temperatures of $q=5$ clock model. 
The scaled variable for the horizontal axis $X(c,t)$ is 
defined by Eq.~(\ref{scaled_variable}).
The estimated values are: (a) $T_1$=\clc\ and (b) $T_2$
=\cld\ for the regular model, and 
(c) $T_1$=\vlc\ and (d) $T_2$=\vld\ for the Villain model.
}
\label{fig_FSS_q5}
\end{figure}

Similar procedure is applied to obtain plot shown in Fig.~\ref{fig_FSS_q5},
which is the FSS plot of Fig.~\ref{fig_R_q5}, associated with
the data for regular and Villain model for $q$=5. 
The estimates of $T_1$ and $T_2$ transition temperatures 
for regular model are respectively \clc\ and \cld. 
The two transitions are very close to each other 
in comparison to the transitions for $q=6$ of the same model. 
This is related to the role played by the discreteness 
which almost rules out the BKT phase.  
For the Villain model, the estimate for $T_1$ and $T_2$ are 
respectively \vlc\ and \vld. 
Here, the temperature range is relatively wider compared to 
that of regular case.
%which is again related to the presence of auxiliary field.

%--------------------------------Fig_FSS_q6------------------------------
%-------------------------------------------------------------------------

%--------------------------------Fig_FSS_q5------------------------------
%-------------------------------------------------------------------------

The obtained set of transition temperatures, $T_1$'s and $T_2$'s, 
are tabulated in Table \ref{TD}, where the duality relation 
(Eq.~(\ref{DL})) are checked for the Villain type. 
As indicated, the duality relation is verified, both for 
$q=6$ and $q=5$, which indicates the accuracy of the present 
calculation. 

\begin{table}
\caption{(Color online) 
Transition temperatures $T_1$ and $T_2$ 
of $q$-state clock models of regular and Villain types
for $q=5, 6$.  For the Villain types, the duality relation, 
Eq.~(\ref{DL}), is checked.}
\label{TD}
\begin{center}
\begin{tabular}{cccccc}
\hline
\hline
$q$& $T_1$ & $\eta(T_1)$ & $T_2$ & $\eta(T_2)$ & $T_1T_2/(4\pi^2/q^2)$ \\
\hline
6 (regular) & \ \cla & \ 0.120 & \ \clb & \ 0.236 & \\
5 (regular) & \ \clc & \ 0.176 & \ \cld & \ 0.234 & \\
6 (Villain) & \ \vla & \ 0.129 & \ \vlb & \ 0.233 & $1.0002$ \\
5 (Villain) & \ \vlc & \ 0.176 & \ \vld & \ 0.235 & $1.0002$ \\
\hline
\end{tabular}
\end{center}
\end{table}

\subsection{FSS without estimating $T_{BKT}$}

We here perform the FSS analysis without estimating $T_{BKT}$ 
to investigate the universality of the BKT transition. 
Such approach was recently employed in the study of 
the generalized XY model \cite{komura}, where the analysis 
for the Potts model \cite{Caraccido,Salas} was applied to 
the study of the BKT transition. 
We focus on the FSS property of the ratio of the correlation ratio 
of different sizes, that is, $R(2L)/R(L)$.

\begin{figure}[t]
\begin{center}
\includegraphics[width=0.48\linewidth]{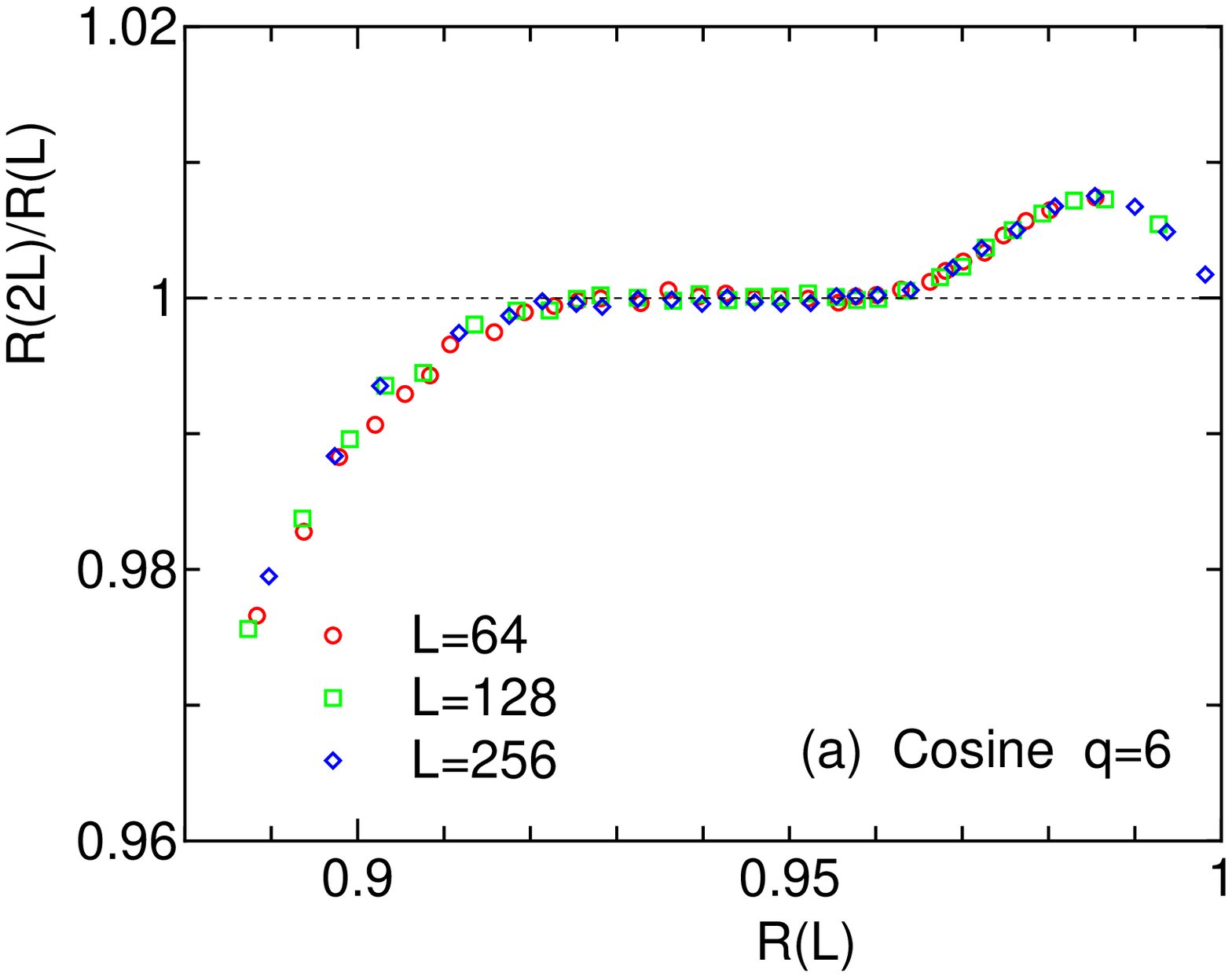}
\includegraphics[width=0.48\linewidth]{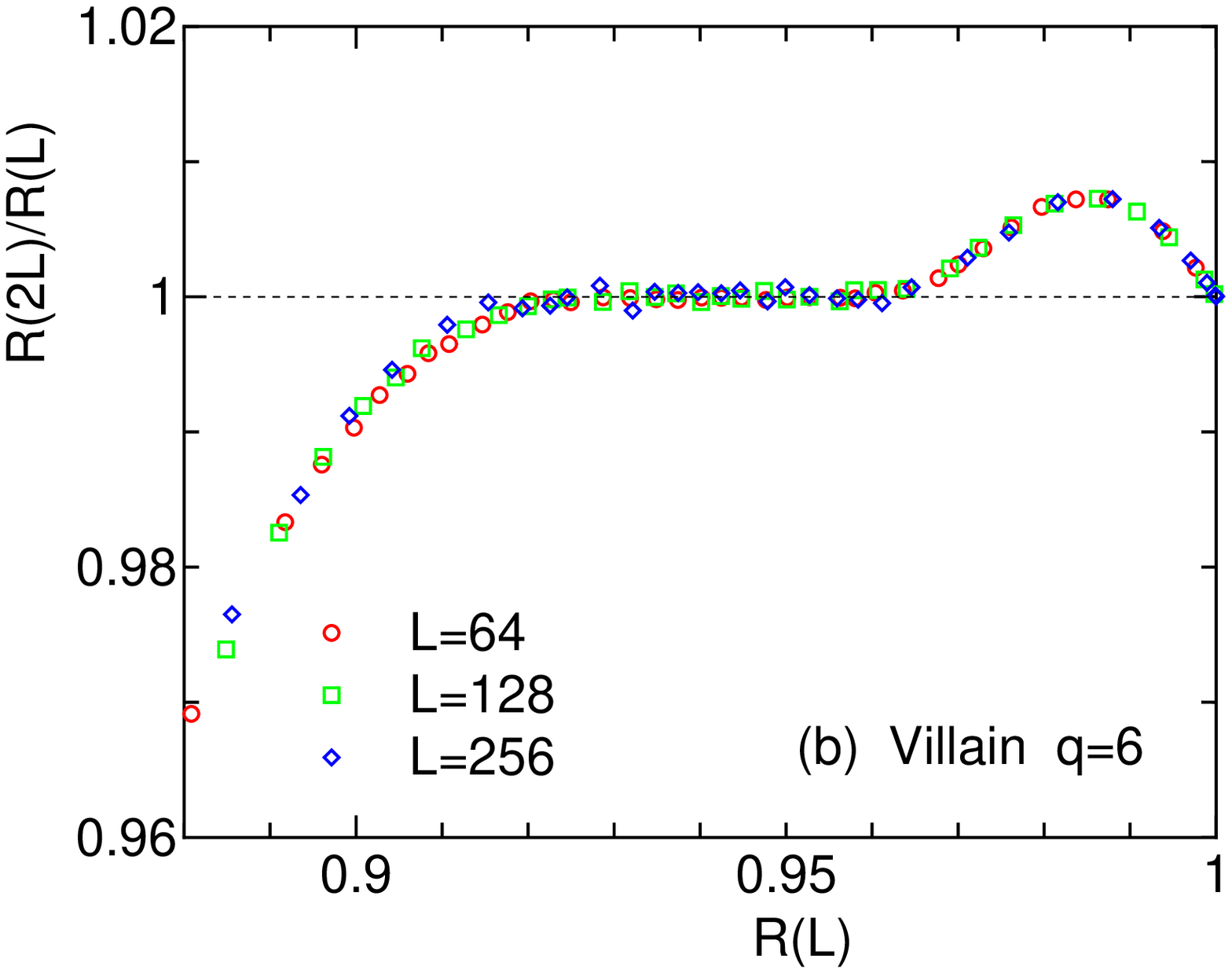}
\includegraphics[width=0.48\linewidth]{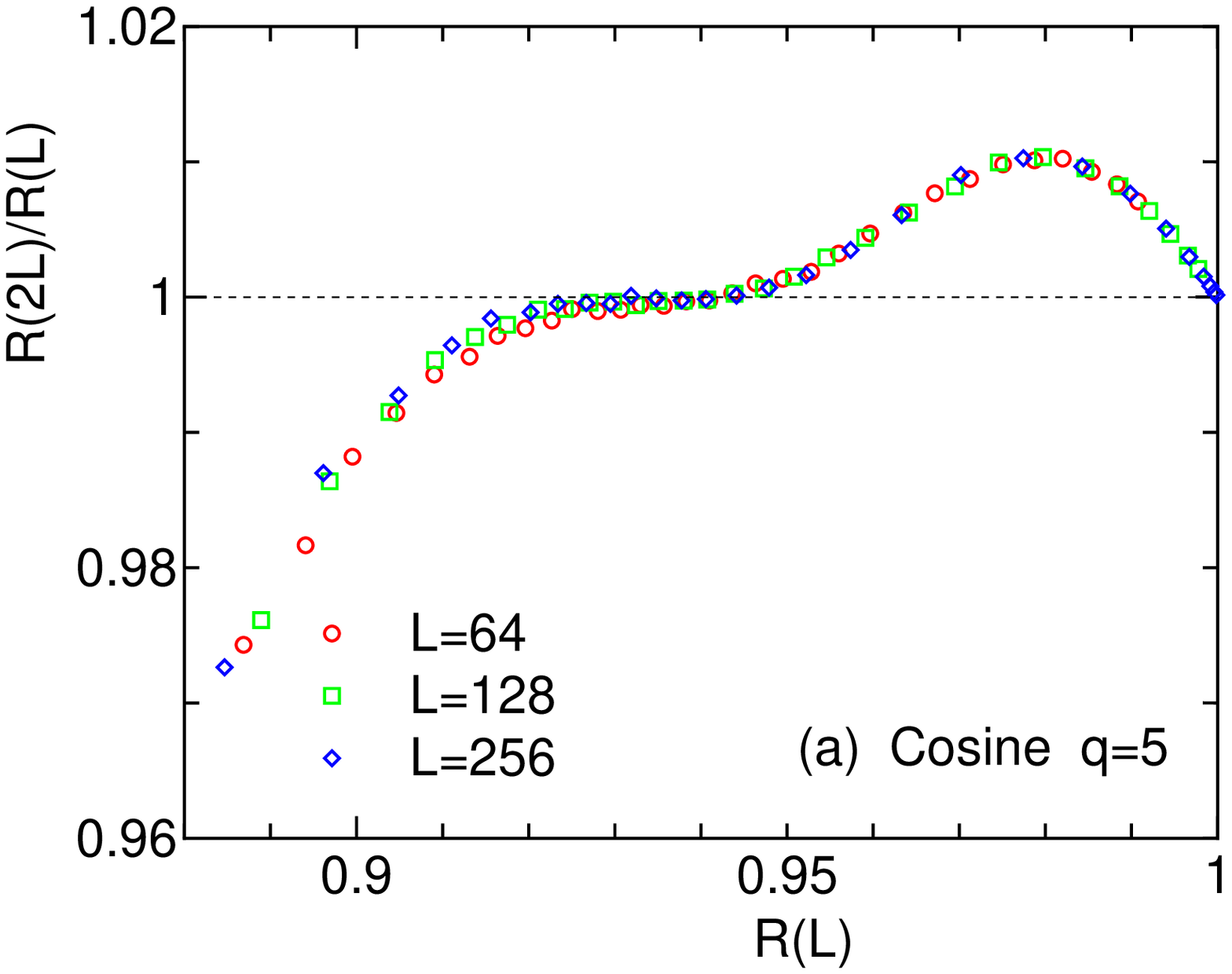}
\includegraphics[width=0.48\linewidth]{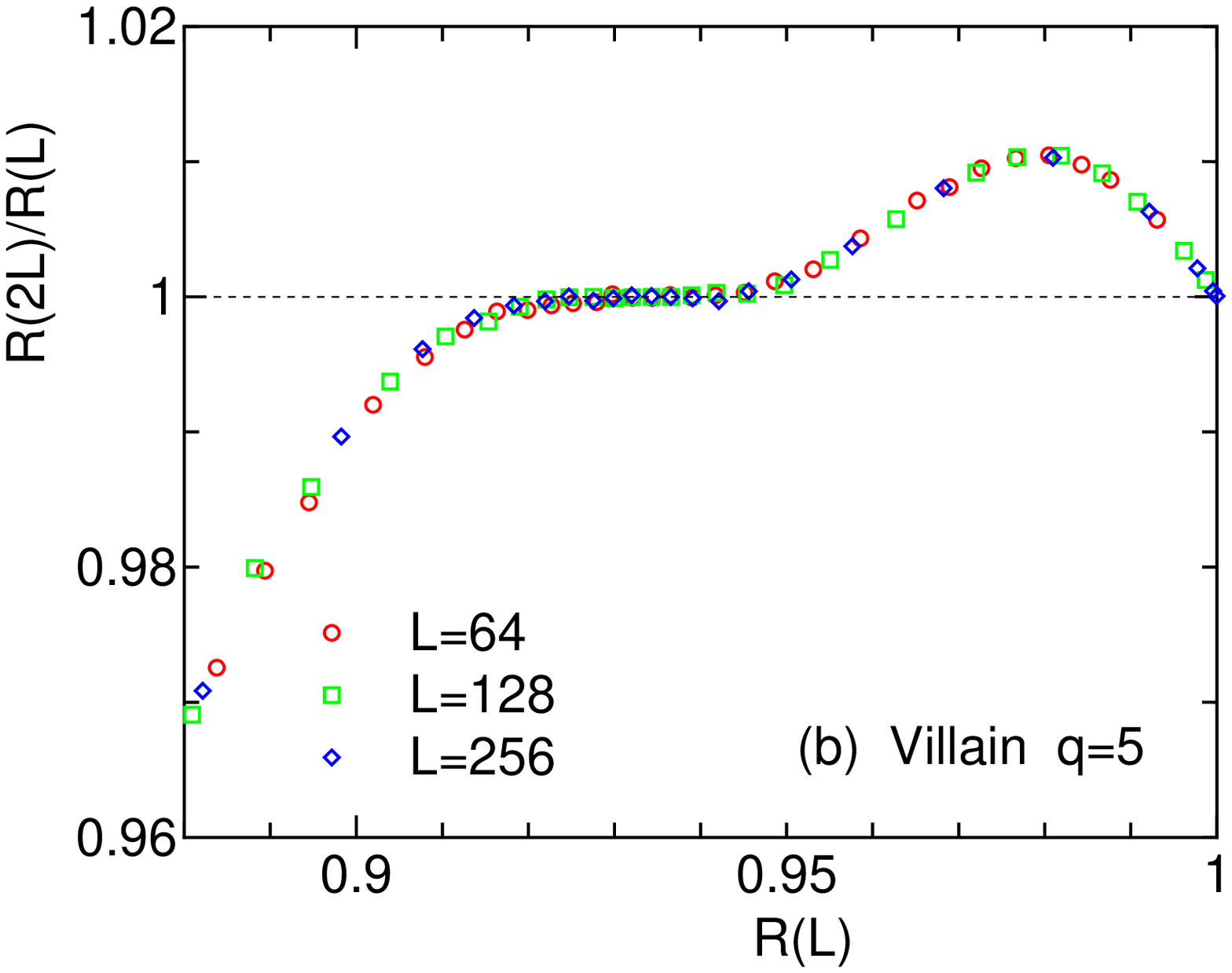}
\end{center}
\caption{(Color online) 
FSS plot of $R(2L)/R(L)$ versus $R(L)$
of (a) regular and (b) Villain types for $q=6$, 
and (c) regular and (d) Villain types for $q=5$. 
The data for the pair of system sizes $L$-$2L$ 
with $L$=64, 128, and 256 are shown.} 
\label{fig_L2L}
\end{figure}

We plot $R(2L)/R(L)$ as a function of $R(L)$ in Fig.~\ref{fig_L2L} 
for four models, (a) regular and (b) Villain models for $q$=6 and 
(c) regular and (d) Villain models for $q$=5. 
The lattice size pairs $L$-$2L$ with $L$=64, 128 and 256 are used. 
All the data for different sizes are collapsed 
on a single curve for each model. That is, the FSS is very good, 
although logarithmic corrections might not be small 
in the BKT transition \cite{kosterlitz2,Janke}. 
The value of $R(2L)/R(L)$ remains 1 for the intermediate BKT phase, 
and becomes smaller than 1 for smaller $R(L)$, 
which means $T > T_2$, as in the XY model \cite{komura}.  
The regime at which the collapsing curves are parallel 
to the horizontal axis with the value of 1 is equivalent to
the collapsing regime of the plot in Fig.~\ref{fig_R_q6} and 
Fig.~\ref{fig_R_q5}, the regime for the existence of BKT phase. 
For clock model of $q=5,6$, the value of $R(2L)/R(L)$ 
becomes larger than 1 for larger $R(L)$, that is, $T < T_1$, 
and comes back to 1 for $R(L)=1$, that is, $T=0$; 
this behavior comes from the ordered phase. 

We assign the values of $R(L)$ at $T_1$ and $T_2$ as 
$R_1$ and $R_2$, respectively.  The values of $R_1$ 
and $R_2$ of regular model and Villain model are 
very close but there is a small difference for $q=6$.  
For $q=5$, the intermediate region becomes narrower, 
and the difference of regular and Villain models 
becomes more appreciable.  The overall FSS behaviors 
of regular and Villain models look the same, but quantitatively 
two are not identical. This difference becomes larger 
for $q=5$, because the difference of the local Boltzmann factor 
becomes larger for high temperatures. 

%--------------------------------fig_L2L----------------------------------
%-------------------------------------------------------------------------

\begin{figure}[t]
\begin{center}
\includegraphics[width=0.48\linewidth]{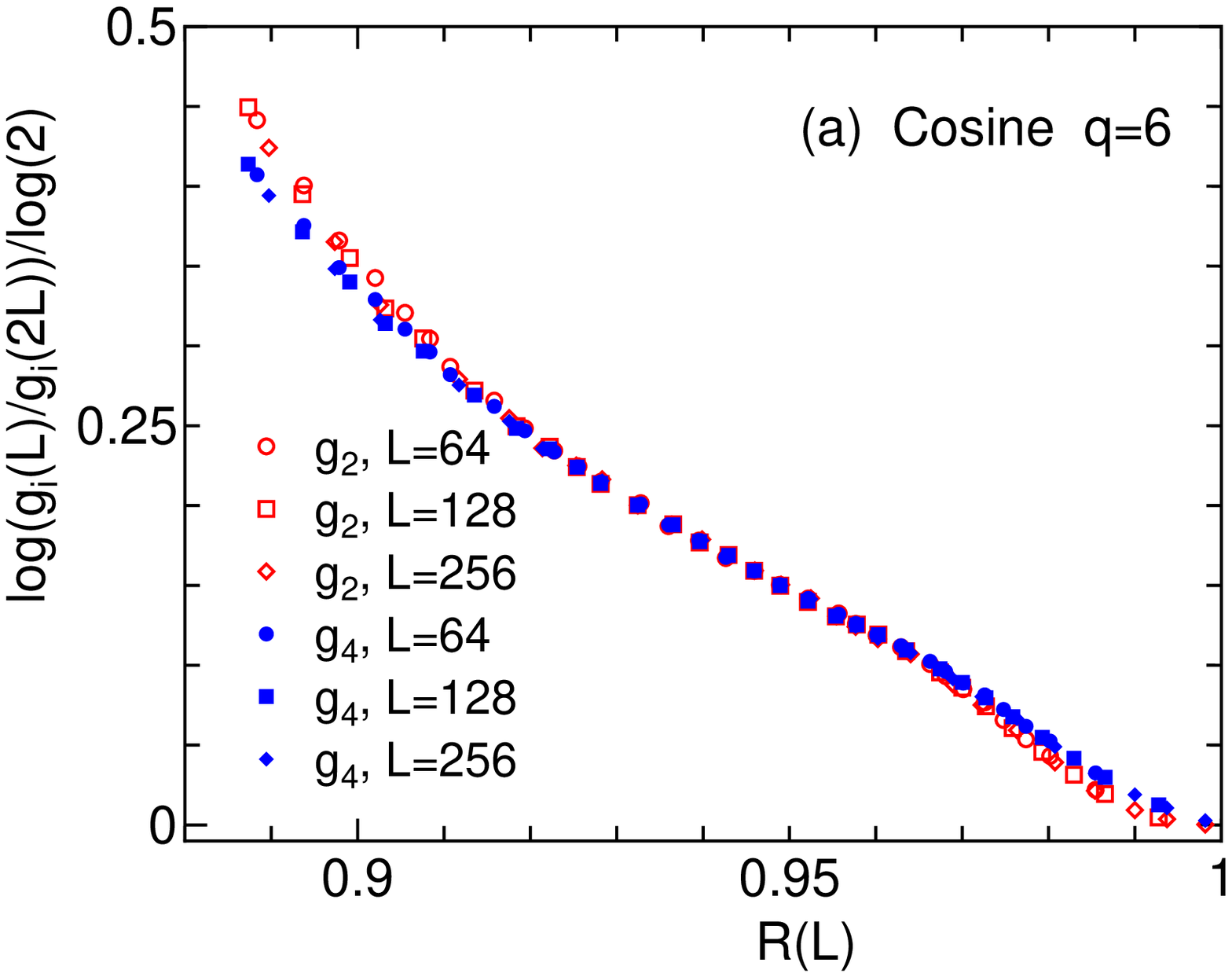}
\includegraphics[width=0.48\linewidth]{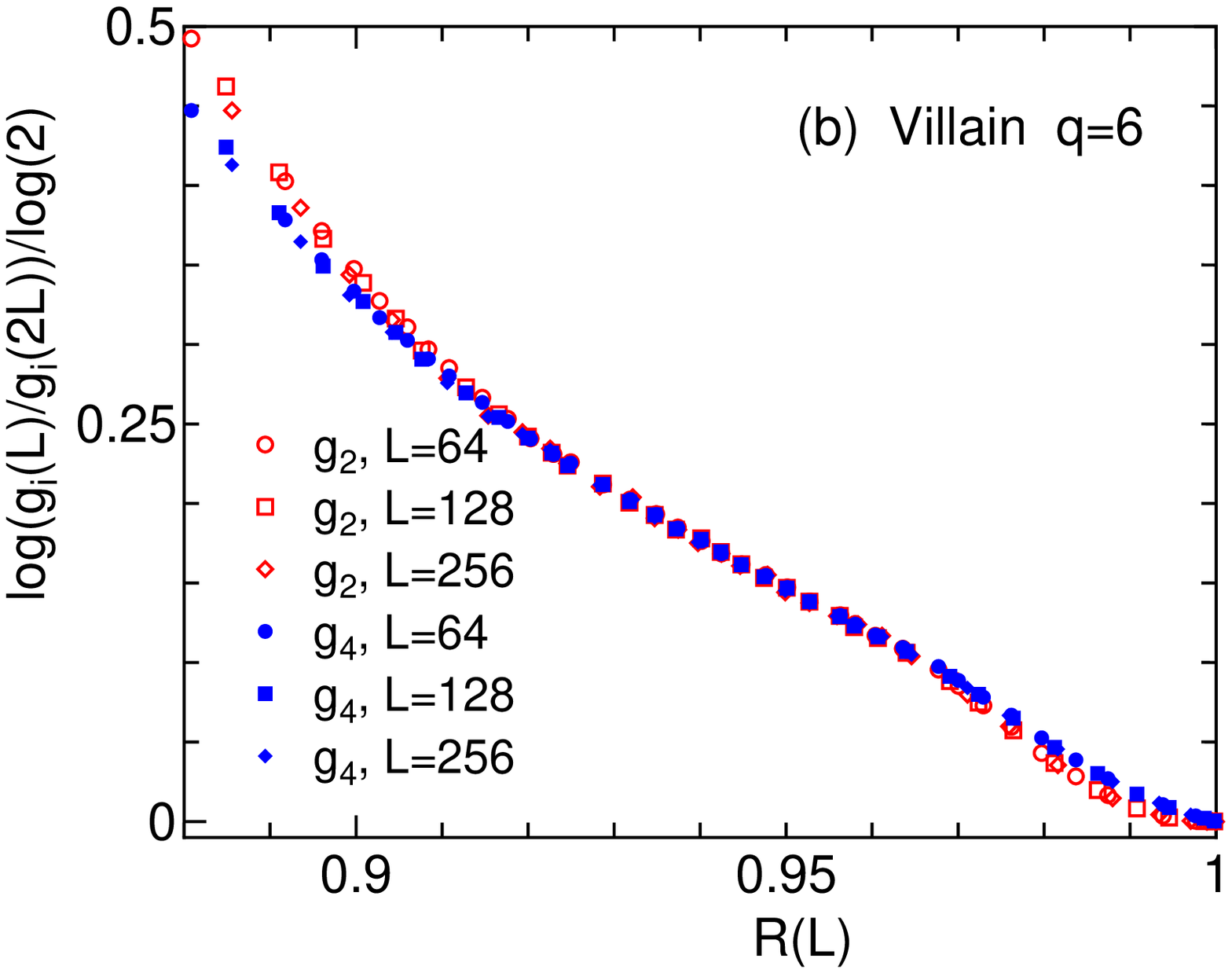}
\includegraphics[width=0.48\linewidth]{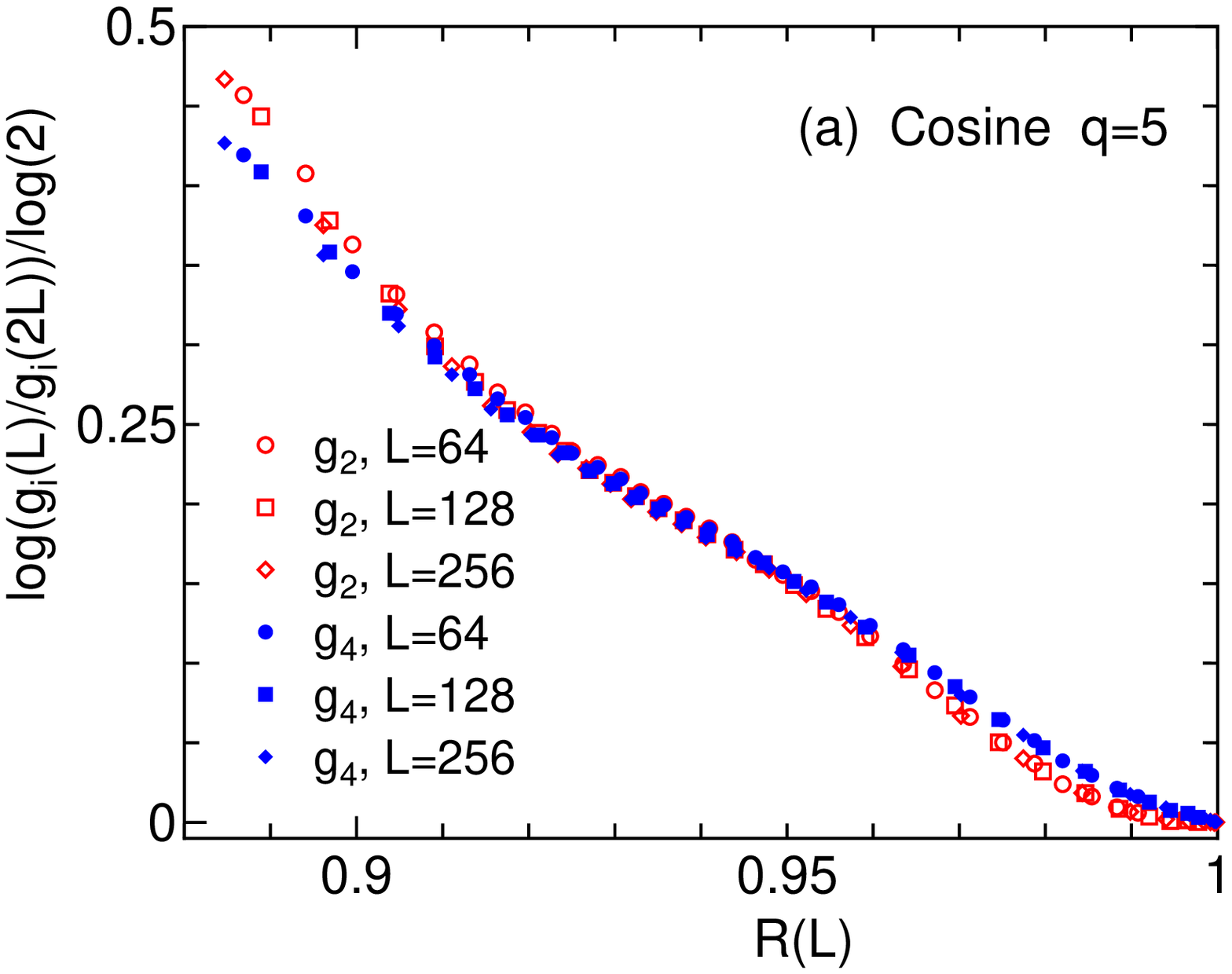}
\includegraphics[width=0.48\linewidth]{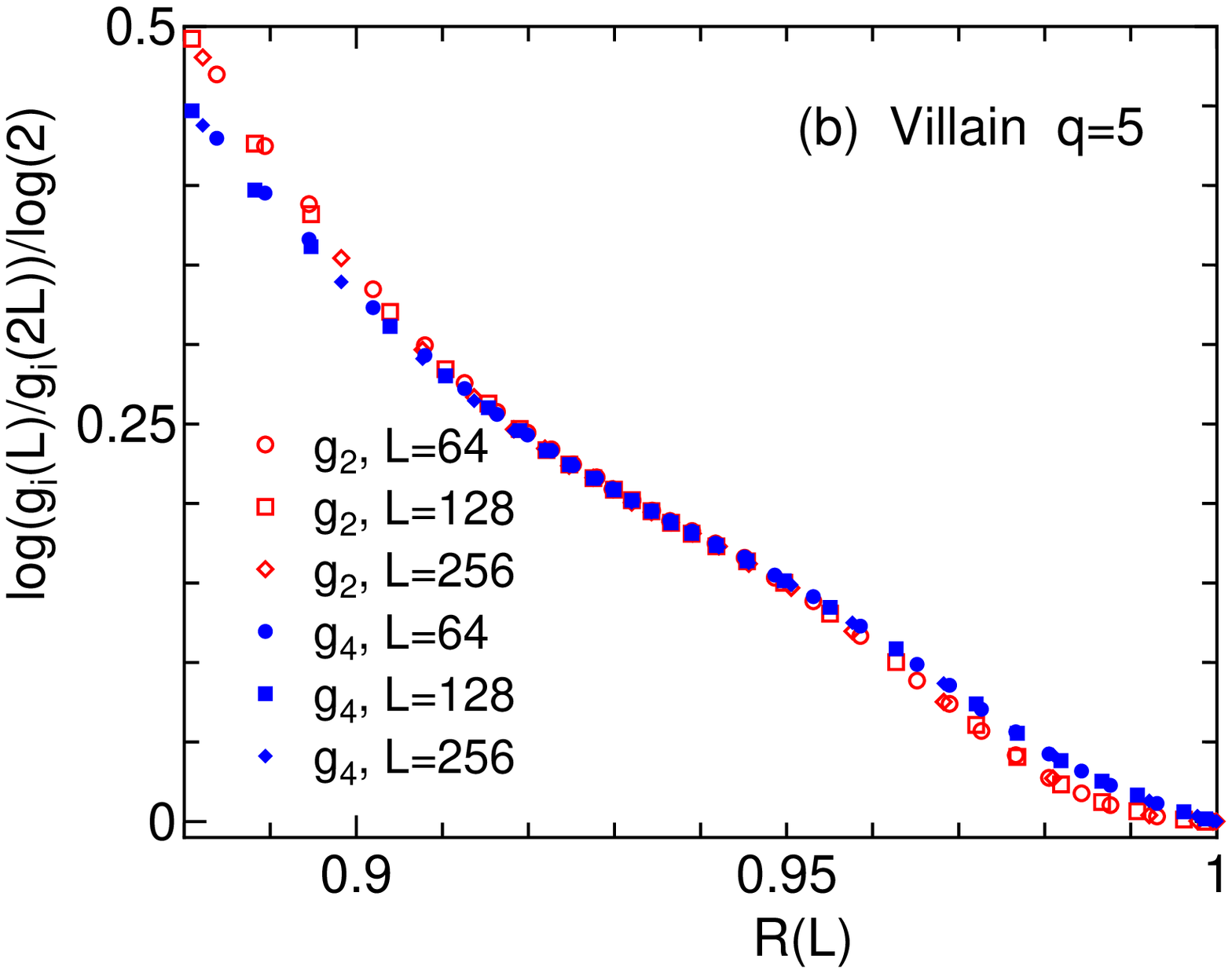}
\end{center}
\caption{(Color online) 
FSS plot for the estimate of decay exponent $\eta$ 
of (a) regular and (b) Villain types for $q=6$, 
and (c) regular and (d) Villain types for $q=5$. 
The value of $\ln [g_2(L)/g_2(2L)]/\ln 2$ gives $\eta$ 
at the temperature of $R(L)$.
} 
\label{fig_eta}
\end{figure}

The temperature dependence of decay exponent $\eta(T)$ is 
extracted from the data of correlation function. 
This is done based on Eq.~(4), $g(L) \sim L^{-\eta(T)}$, 
from which we obtain the following scaling form,
\begin{equation}
  \frac{g_2(L)}{g_2(2L)}= \frac{L^{-\eta}f(L/\xi) }
  {(2L)^{-\eta}f(2L/\xi)}=2^{\eta}\frac{f(L/\xi)}{f(2L/\xi)}
\end{equation}
with $g_i(L) = g(L/i)$ for the system with linear size $L$.
By plotting $\ln [g_2(L)/g_2(2L)]/\ln 2$ 
with respect to the correlation ratio $R(L)=g_2(L)/g_4(L)$, 
as shown in Fig.~\ref{fig_eta}, we can obtain the value of $\eta$ 
at lower and higher BKT temperatures. The horizontal axis $R(L)$ 
can effectively act as temperature axis. Using the estimated 
values of $R_1$ and $R_2$, we estimate 
$\eta(T_1)=0.120$ and $\eta(T_2)=0.236$ for regular model of $q=6$.
With this procedure, $\eta(T_1)$ and $\eta(T_2)$ are estimated 
for regular and Villain models with $q=6$ and $q=5$. 
The estimated exponents are tabulated in Table \ref{TD}. 
We note that the estimated $\eta(T_2)$'s are a little bit 
smaller than the theoretical value of 1/4.  At the same time, 
the estimated $\eta(T_1)$'s are a little bit larger than 
the theoretical value of $4/q^2$, that is, 1/9=0.1111 for $q=6$ 
and 4/25=0.16 for $q=5$.  This small systematical deviation 
is attributed to the logarithmic corrections \cite{kosterlitz2, Janke}. 

%--------------------------------Fig_eta------------------------------
%-------------------------------------------------------------------------

\section{Summary and Discussions}

%The correlation ratio shows the symmetrical behavior 
%of lower side and lower side, implying that 
%the existing two transitions should belong to the same universality.
%Transforming using duality emphasis that that was the case,
%no difference between the two. 

We have investigated 2D $q$-state clock models of regular (cosine) 
and Villain type for $q=5,6$ by using the large-scale Monte Carlo 
simulations.  By calculating the correlation ratio 
and the size-dependent correlation length for both types 
for each value of $q$, we observed double transitions at each model.
The range of temperature in which BKT phase exists is
wider for $q=6$ than for $q=5$, for both regular
and Villain models. This is an indication that
for $q=5$, the discreteness starts to play a major
role in creating true LRO at lower temperature,
although not yet totally rules out the 
existence of BKT phase.

Verifying the duality relation for the Villain type, 
and confirming the universal behavior for regular and Villain types, 
we conclude that the $q$-state clock models have 
double transitions of BKT type for $q=5,6$.  
As for the recent controversy, our result coincides with 
that of Kumano {\it et al.} \cite{yuta} and that of 
Chatelain \cite{crist}. 

Our conclusion disagrees with those by some references 
\cite{lapilli,hwang,baek02}.  In Ref.~\cite{baek02}, 
%they used the helicity modulus in characterizing the BKT transition.  
they used  helicity modulus defined with respect to
an infinitesimal twist, which is inappropriate when characterizing
the BKT transition of the clock models.  Although the helicity modulus 
has been successfully used in the study of the BKT transition, 
it fails to characterize the lower transition. 
We briefly comment on the helicity modulus in Appendix A. 

In the case of $q$=4, the clock model becomes equivalent to 
two sets of Ising model, and exhibits 2nd-order transition. 
It is instructive to show the present analysis in the case 
of the 2nd-order transition.  We show the data of $q=4$ 
for both regular and Villain models in Appendix B. 

To summarize, we have shown that $q$-state clock models 
exhibit double BKT transitions for regular and Villain types 
for $q>4$, although the region of BKT phase is narrow 
for $q=5$ regular clock model, and the corrections to scaling 
is large for this case.  It will be interesting to 
see the phase diagram of the clock model, that is, 
the plot of transition temperatures as a function of $q$, 
which is given in Appendix C.

%%%%%%%%%%%%%%%%%%%%%%%%%%%%%%%%%%%%%%%%%%%%%%%%%%%%%%%%%%%%%%%

\section*{Acknowledgments}

The authors wish to thank Toshihiko Takayama for valuable discussions.  
This work was supported by a Grant-in-Aid for Scientific Research 
from the Japan Society for the Promotion of Science. 
TS is grateful to Japan Student Service Organization (JASSO) 
for its hospitality and to KLN research grant of Hasanuddin University,
FY 2018.

\clearpage

\appendix

%--------------------------------fig_helicity-----------------------------
\begin{figure}[t]
\includegraphics[width=0.48\linewidth]{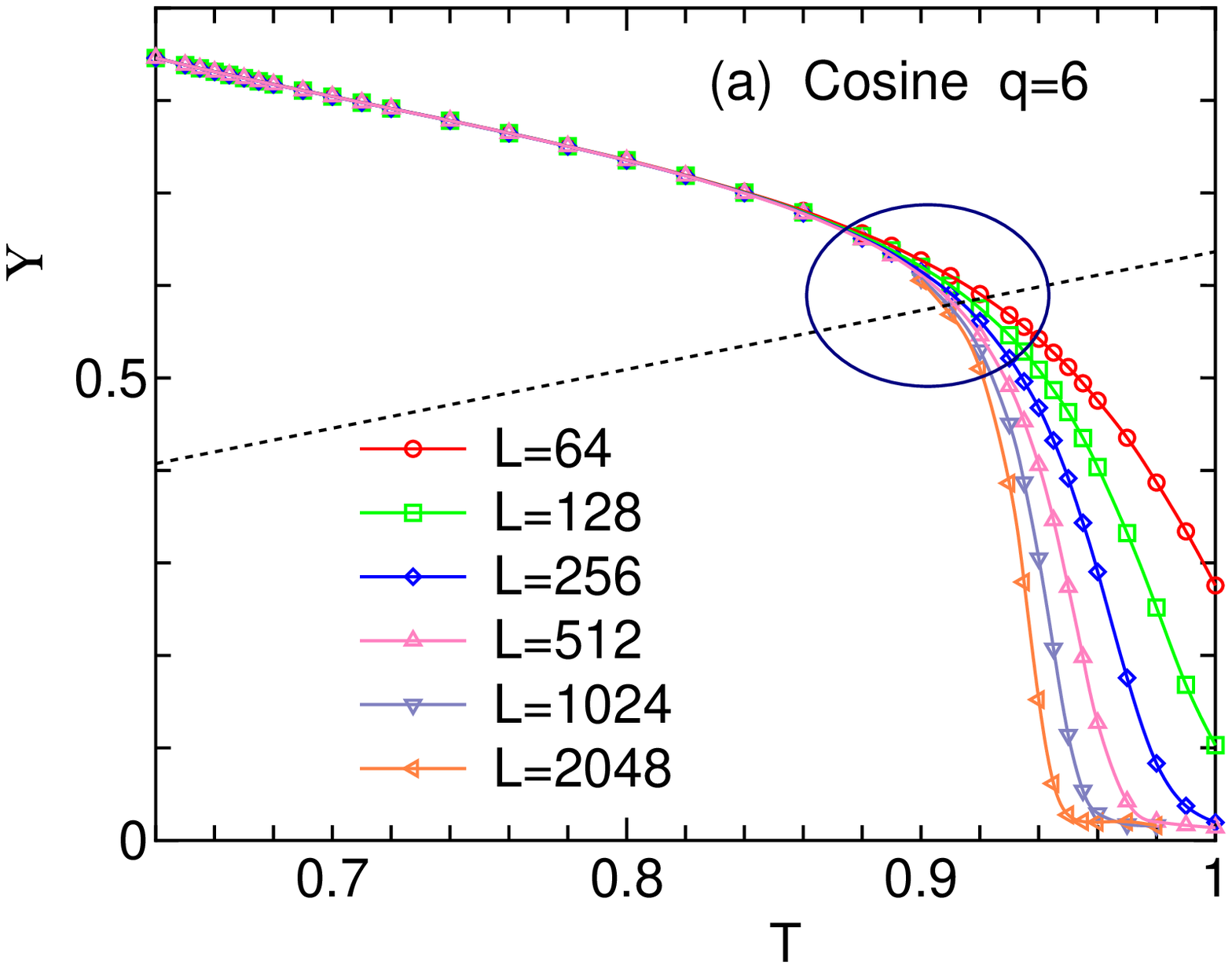}
\includegraphics[width=0.48\linewidth]{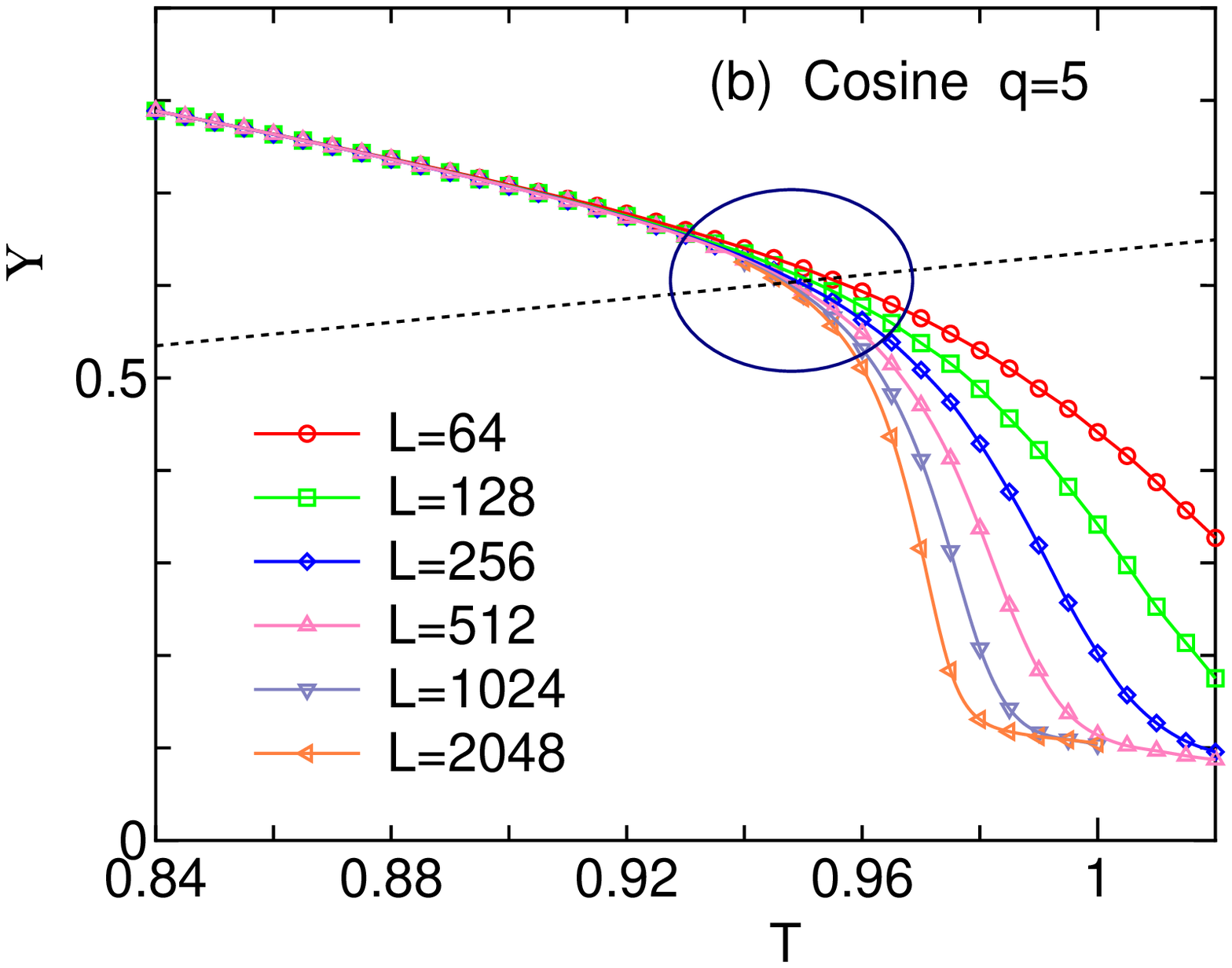}
\caption{(Color online) 
The helicity modulus $\Upsilon$ as a function of temperature 
for the regular model with (a) $q=6$ and (b) $q=5$. 
The universal jump is expected at the point where $\Upsilon=2T/\pi$.}
\label{fig_helicity}
\end{figure}
%-------------------------------------------------------------------------

\section{Helicity modulus}

The helicity modulus  can be expressed as
\begin{equation}
 \Upsilon = <e> - \frac{L^2}{T} <s^2>
\end{equation}
where
\begin{equation}
 e = \frac{1}{L^2} \sum_{<ij>_x} \cos(\theta_i - \theta_j),
\end{equation}
\begin{equation}
 s = \frac{1}{L^2} \sum_{<ij>_x} \sin(\theta_i - \theta_j),
\end{equation}
and the sum is over all links in one direction.

The helicity modulus $\Upsilon$ is plotted in Fig.~\ref{fig_helicity} 
as a function of temperature for the regular model 
with $q=6$ and $q=5$. 
The universal jump is expected at the point where $\Upsilon=2T/\pi$. 
The estimate of $T_2$ by the assumption \cite{weber,harada}
\begin{equation}
 \frac{\pi \Upsilon}{2T} \propto 1 + \frac{1}{2 \ln[L/L_0(T)]}
\end{equation}
gives consistent values with the estimates by FSS analysis 
tabulated in Table \ref{TD}. 
There are no singularities at $T_1$, which means that the helicity modulus
is not a good estimator to probe the lower BKT transition.

%The authors of Ref.~\cite{baek02} claimed that for $q=5$ of regular type 
%the helicity modulus does not become zero quickly 
%just above the KT temperature.  It comes from large corrections to scaling, 
%and it does not mean that the transition is not KT type. 
%
The authors of Ref.~\cite{baek02} claimed that
the transition for $5$-state clock model is not BKT type as
the helicity modulus does not vanish
 above temperature transition.  
%We argue that this is  due to the use of the incommensurate
%form of Helicity modulus for clock model \cite{yuta},
%and therefore the remnant does not mean that the transition is not BKT type. 
We argue that this is due to the improper way of calculating
helicity modulus for the clock model, and therefore the remnant
does not mean that the transition is not KT type.  The correct
way of calculating helicity modulus for the clock model has to
take into account the existing $Z_q$ discrete symmetry, which
is done by a finite, quantized twist as  has been described by
Kumano et al. \cite{yuta}, rather than the infinitesimal twist \cite{baek02}.
As well known, in the BKT phase, spin configurations are dominated by
the vortex-antivortex pairs.  The number of  pairs in the BKT phase for
the continuous XY model can be very large as spins are allowed to change orientation
continuously. The situation is different for the case of $q$-state clock model
where the smaller the value of $q$, the smaller the possible number of
pairs in BKT phase. In the case of $q=5$, which is just above the critical
value, $q=4$, the occurrence of vortex-antivortex unbinding
can not be precisely detected in relatively small system sizes;
as any possible change of spin orientation is always large.

The critical anomaly associated with the higher BKT transition appears 
near the temperature where $\Upsilon=2T/\pi$; this region is encircled 
in Fig.~\ref{fig_helicity}. 
We note that the tiny correction to the critical value of $2/\pi$ 
due to winding field for periodic boundary conditions was 
calculated by Hasenbusch \cite{Hasenbusch2005}, but the difference 
is less than 0.02\%. 
From Fig.~\ref{fig_helicity}(b), we see that the helicity modulus 
remains around 0.1 even for large enough size $L=2048$ 
above $T_2$, say at $T=0.98$. 
The size dependence is very small.  
This comes from the corrections to scaling.  
We also see a remnant of 
such behavior for $q=6$, which is shown in Fig.~\ref{fig_helicity}(a); 
the helicity modulus remains 0.02, say at $T=0.95$, for large size $L=2048$.  
To eliminate this remnant, one needs to modify the form 
of Helicity modulus when probing the BKT phase of clock models as has 
been described by Kumano {\it et al.} \cite{yuta}.

%--------------------------------fig_R_q4--------------------------------
\begin{figure}[t]
\includegraphics[width=0.48\linewidth]{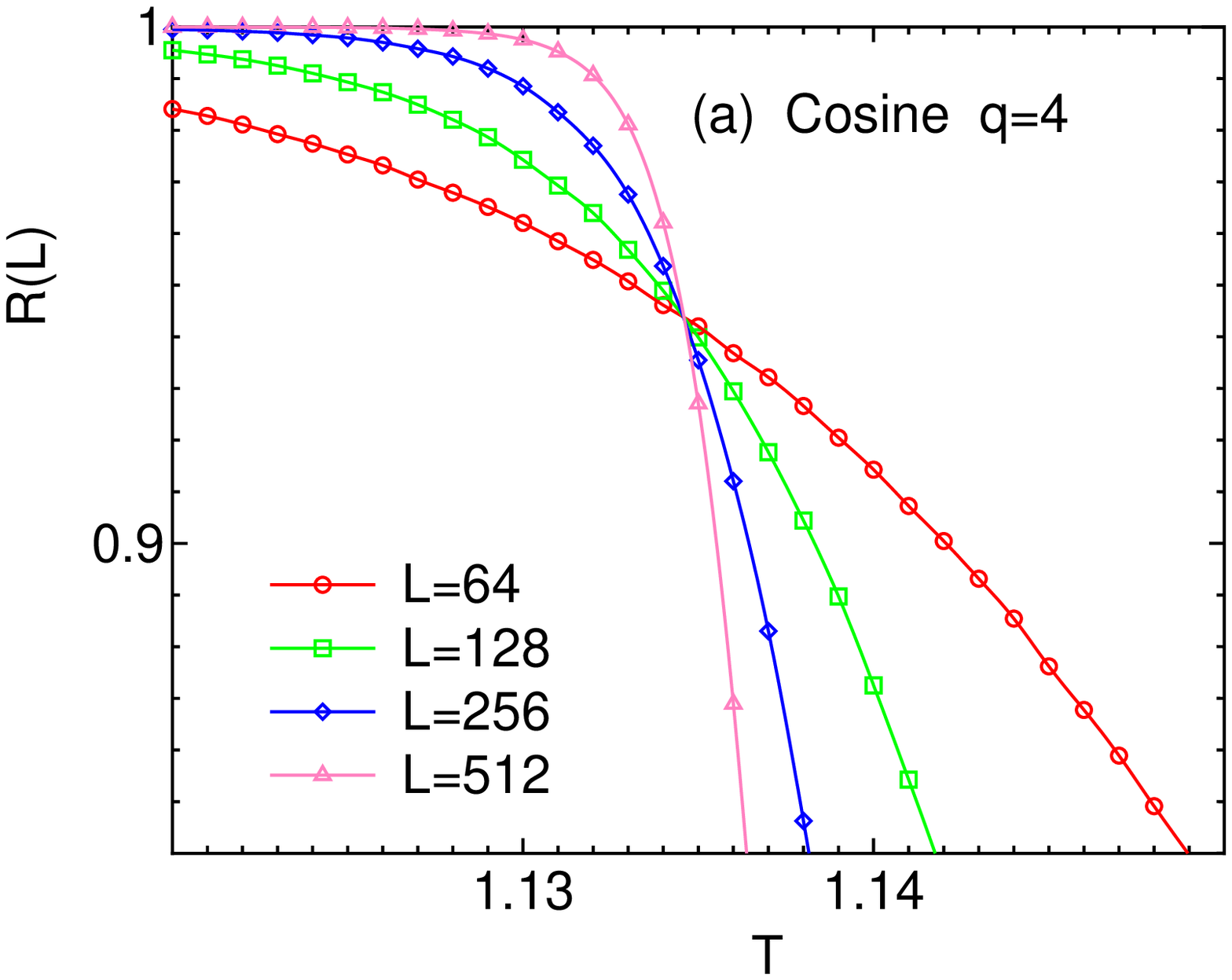}
\includegraphics[width=0.48\linewidth]{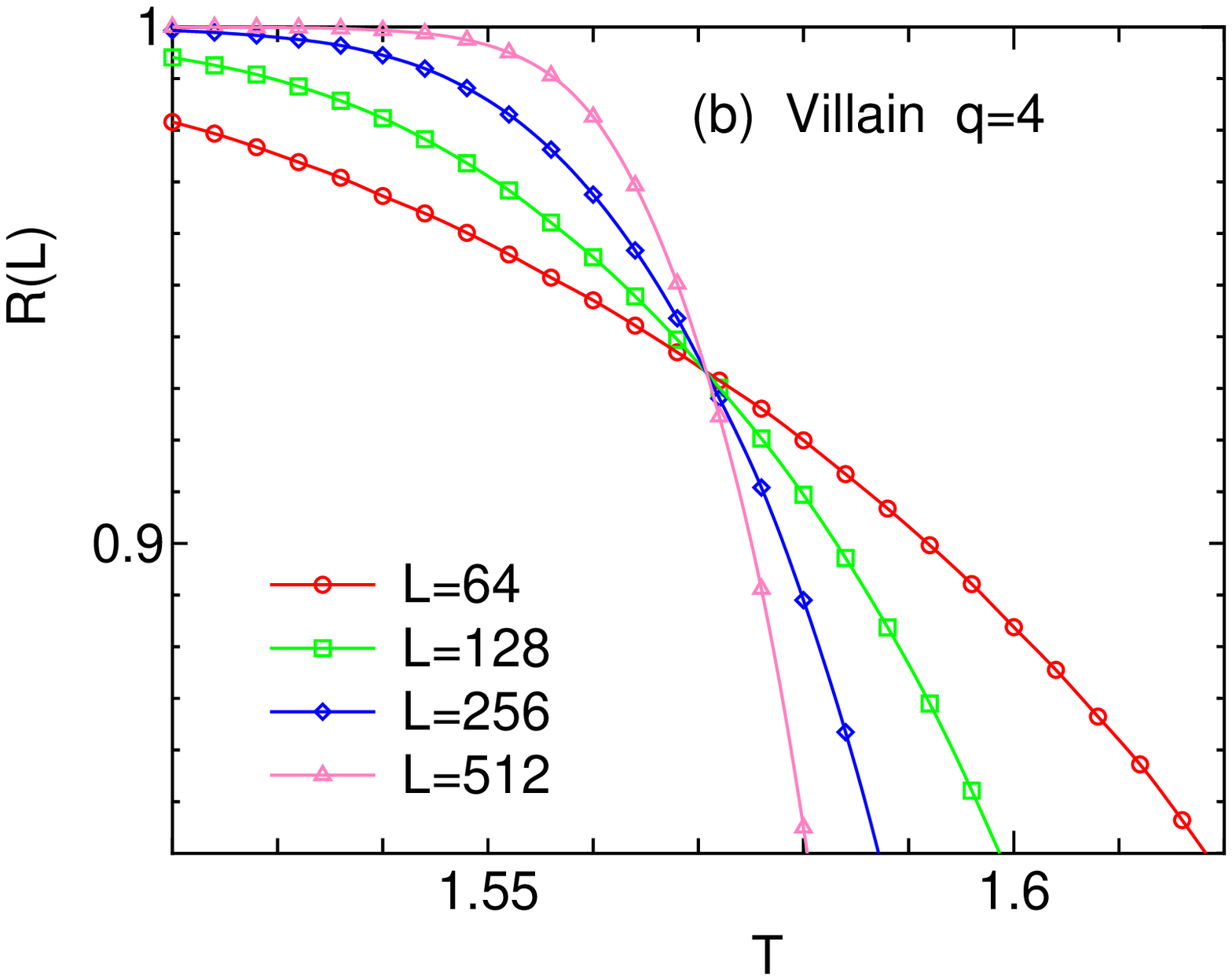}
\caption{(Color online) 
The temperature dependence of correlation ratio, 
$R(L)=g(L/2)/g(L/4)$, for (a) regular and (b) Villain types for $q=4$.  
The crossing of the data with different system sizes gives 
the 2nd-order transition temperature $T_c$. 
}
\label{fig_R_q4}
\end{figure}
%-------------------------------------------------------------------------

%--------------------------------fig_L2L_q4-------------------------------
\begin{figure}[t]
\includegraphics[width=0.48\linewidth]{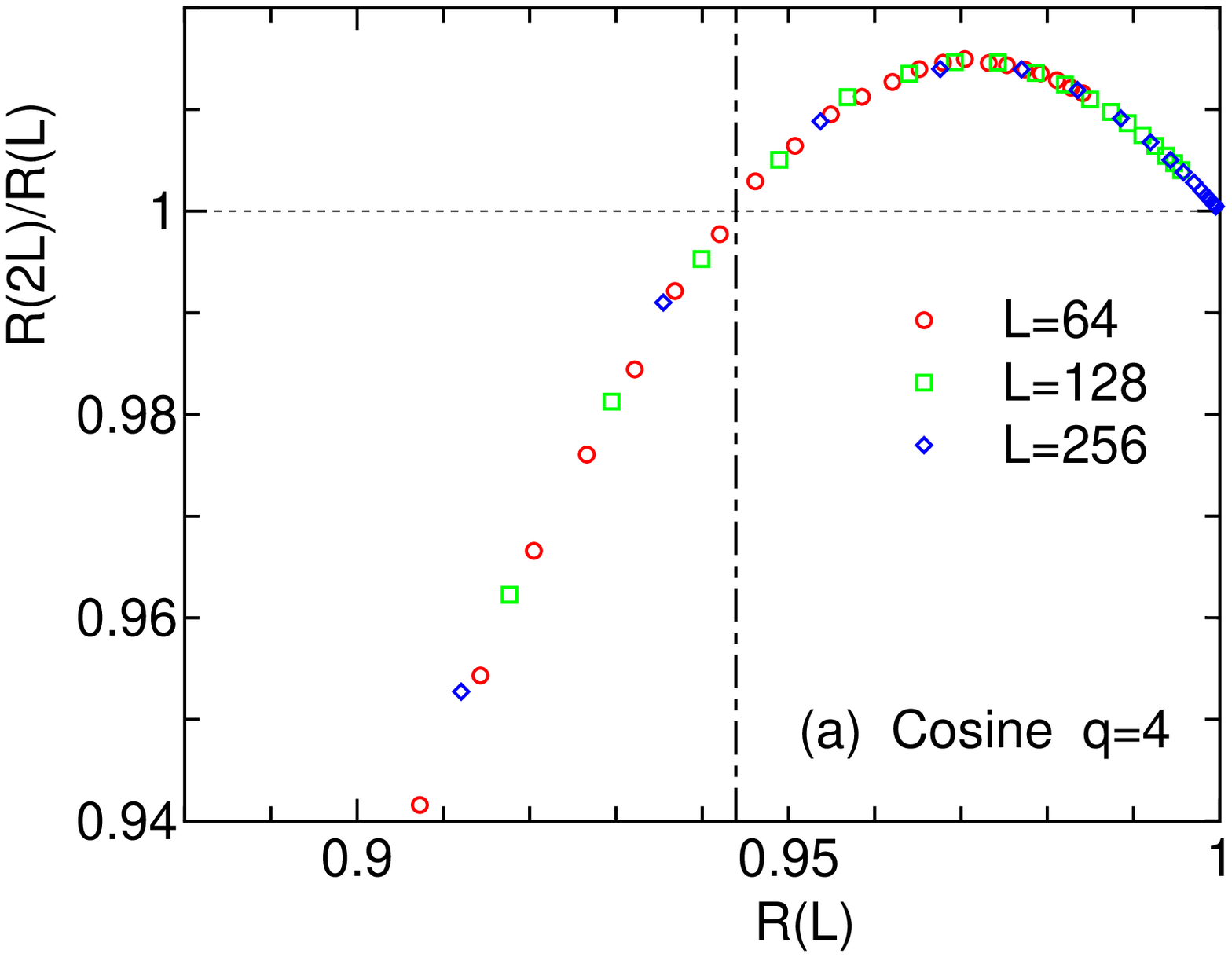}
\includegraphics[width=0.48\linewidth]{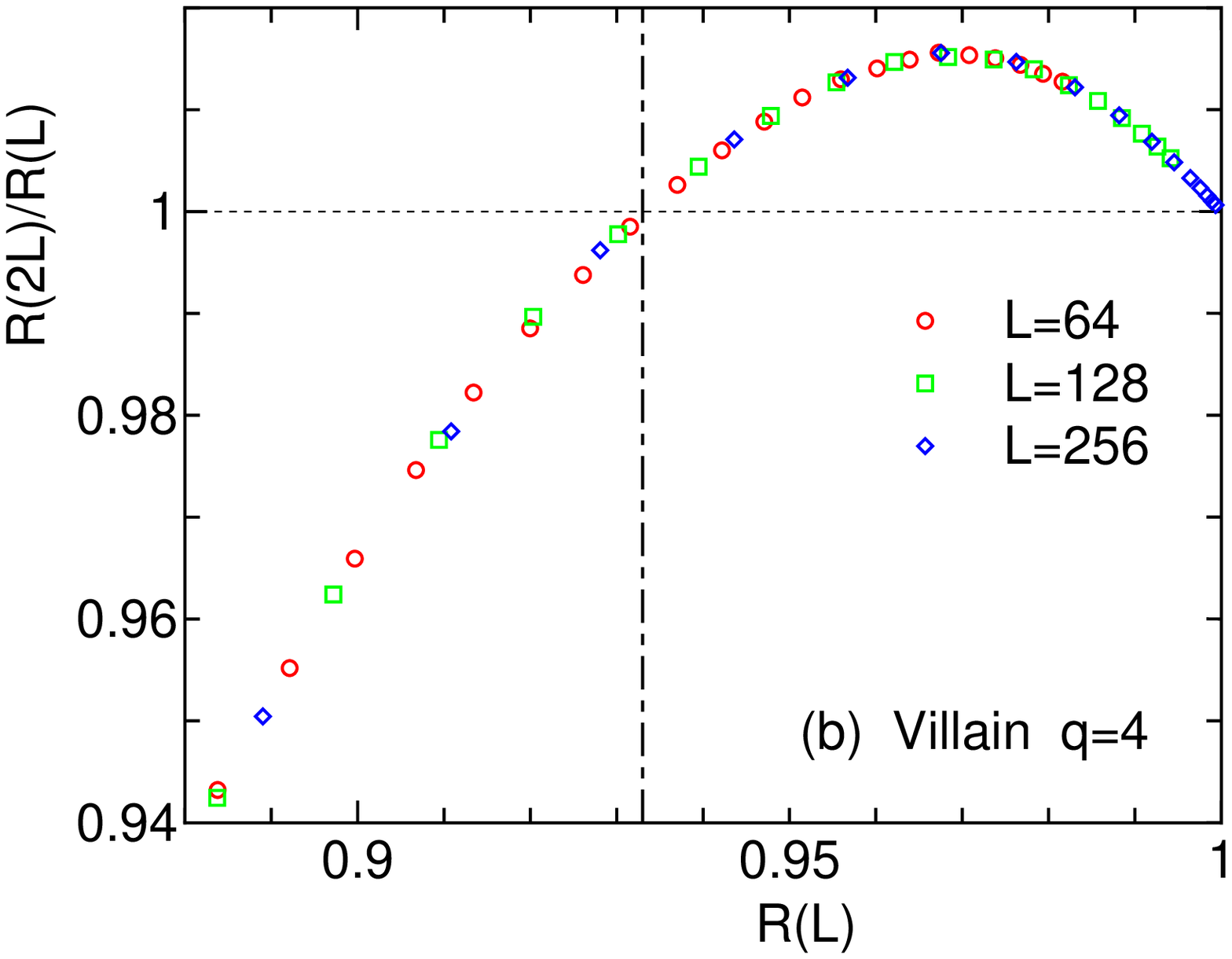}
\caption{(Color online) 
FSS plot of $R(2L)/R(L)$ versus $R(L)$ 
of (a) regular and (b) Villain types for $q=4$.  
The data for the pair of system sizes $L$-$2L$ 
with $L$=64, 128, and 256 are given. 
The value of $R_c$ is shown by the vertical dot-dashed line, 
for convenience.
}
\label{fig_L2L_q4}
\end{figure}
%-------------------------------------------------------------------------
%--------------------------------Fig_eta_q4------------------------------
\begin{figure}[t]
\includegraphics[width=0.48\linewidth]{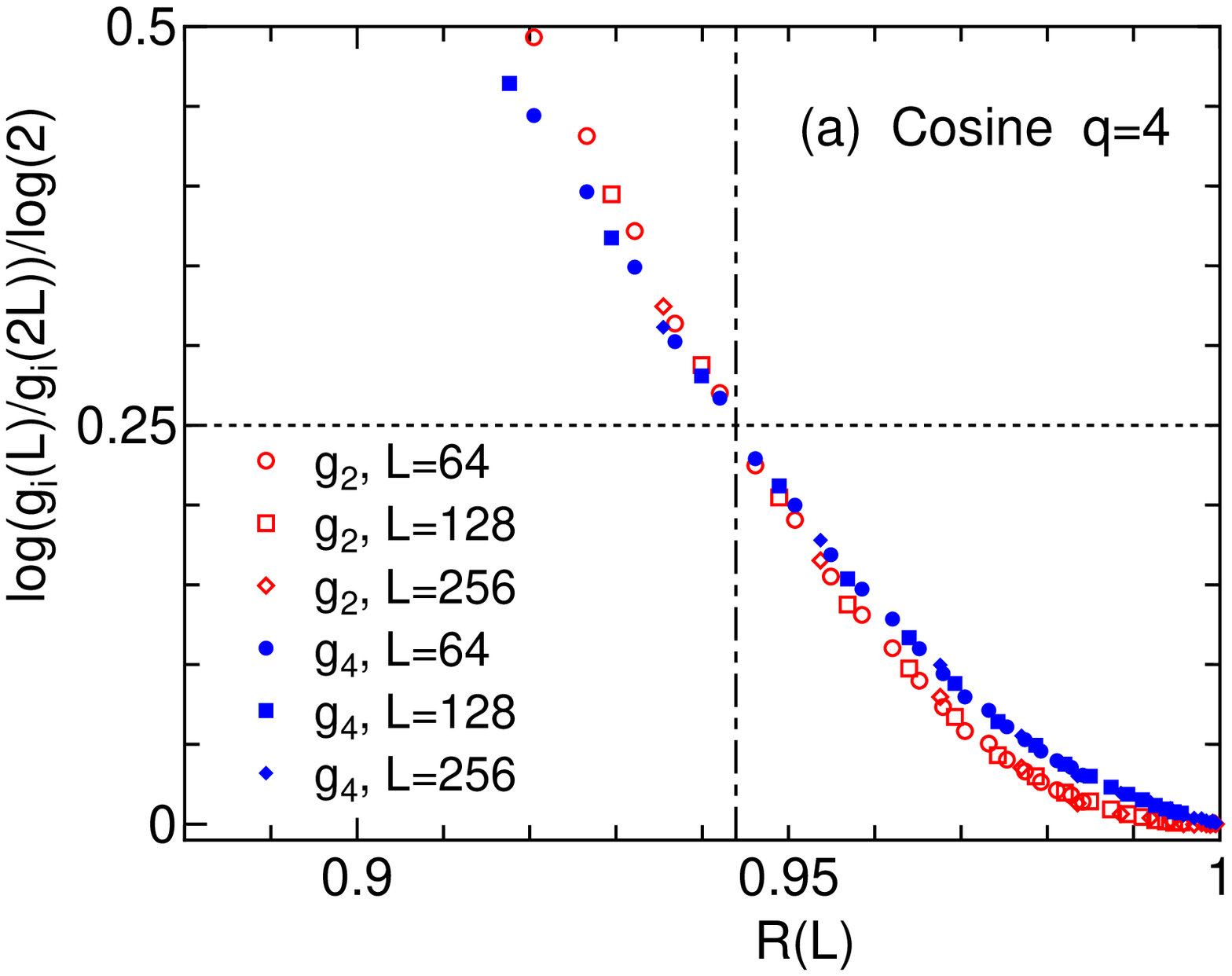}
\includegraphics[width=0.48\linewidth]{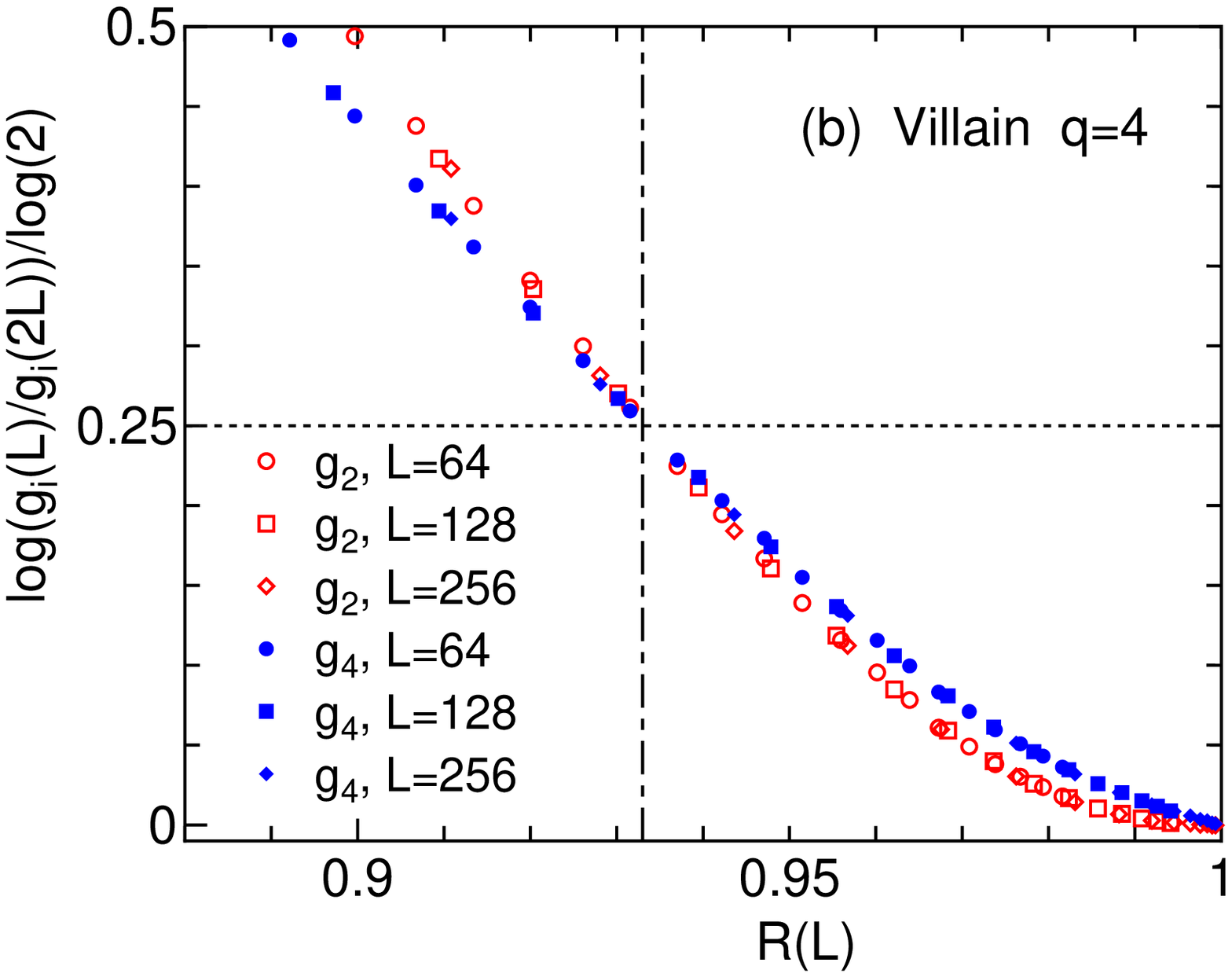}
\caption{(Color online) 
FSS plot for the estimate of decay exponent $\eta$ 
of (a) regular and (b) Villain types for $q=4$. 
The value of $\ln [g_i(L)/g_i(2L)]/\ln 2$ with $i$ = 2, 4 
at the critical value of $R(L)$, $R_c$, gives the estimate of 
$\eta$.  The vertical dot-dashed line shows the value of $R_c$. 
} 
\label{fig_eta_q4}
\end{figure}
%-------------------------------------------------------------------------

\section{Data for $q$=4}

In the case of $q$=4, the clock model becomes equivalent to 
two sets of Ising model, and exhibits 2nd-order transition. 
We show the data of $q$=4 for both regular and Villain models 
for convenience.  The correlation ratios $R(L)$ of regular and 
Villain models for $q=4$ are given in Fig.~\ref{fig_R_q4}. 
We observe the crossing of the data at the 2nd-order 
transition points.  For the regular model, $T_c$ is 
exactly given by $1/\ln(1+\sqrt{2}) = 1.1346$, and 
for the Villain model, $T_c$ is given by $\pi/2 = 1.5708$. 
The estimates of $T_c$ from the crossing of the data 
reproduce the exact values well. 
The expression of the critical value $R_c$ for the regular model 
can be given in terms of the Jacobi $\theta$ functions \cite{Francesco}, 
and the exact numerical value becomes $0.943905 \cdots$ \cite{tomita02b}.
It is to be noted that the estimated critical value $R_c$ for the 
Villain model, 0.933, is different from the value of the regular model. 

The plot of $R(2L)/R(L)$ vs. $R(L)$ for $L$=64, 128, and 256 
is given in Fig.~\ref{fig_L2L_q4}.  We observe a very good FSS. 
The value of $R(2L)/R(L)$ becomes 1 at a single point, $R_c$, 
which indicates the 2nd-order transition.  It differs from 
the situation of the BKT transition, where the flat region 
in this plot, the BKT phase, exists. 
Since the critical value $R_c$ is different for the regular 
and the Villain models, we can say that the FSS function 
for $R(2L)/R(L)$ vs. $R(L)$ is not universal. 

We plot $\ln [g_i(L)/g_i(2L)]/\ln 2$ with $i$ = 2, 4 
in Fig.~\ref{fig_eta_q4}.  We use the (red) open marks 
for $g_2$ and the (blue) closed marks for $g_4$. 
We find that the estimate of $\eta$ from the value at $R_c$ 
is consistent with the exact value 1/4 for both regular 
and Villain models. 
We also note that the data with $i$=2 and those with $i$=4 
cross at the critical value $R_c$. 
It means that we can estimate $\eta$ from this plot 
even if we do not know $T_c$ or $R_c$. 

\begin{table}
\caption{(Color online) 
Transition temperatures $T_1$ and $T_2$ 
of $q$-state clock models of regular and Villain types 
for $q$=4, 5, 6, 8, and 12.  a) Present study. 
b) Ref.~\cite{tomita02}. c) Ref.~\cite{takayama}. 
d) Ref.~\cite{hasenbusch}. 
For $q$=4, the second-order critical temperatures ($T_1=T_2$) 
are exact.
}
\label{TC}
\begin{center}
\begin{tabular}{ccccc}
\hline
\hline
 & \multicolumn{2}{c}{regular model} & \multicolumn{2}{c}{Villain model}\\
$q$ & $T_1$ & $T_2$ & $T_1$ & $T_2$ \\
\hline
4  & \ 1.1346 & \ 1.1346 & \ 1.5708 & \ 1.5708 \\
5  & \ \clc$^{\rm a)}$ & \ \cld$^{\rm a)}$ & \ \vlc$^{\rm a)}$ & \ \vld$^{\rm a)}$ \\
6  & \ \cla$^{\rm a)}$ & \ \clb$^{\rm a)}$ & \ \vla$^{\rm a)}$ & \ \vlb$^{\rm a)}$ \\
6  & \ 0.7014(11)$^{\rm b)}$ & \ 0.9008(6)$^{\rm b)}$ & \ 0.8266(18)$^{\rm c)}$ & \ 1.3269(12)$^{\rm c)}$ \\
8  & \ 0.4259(4)$^{\rm b)}$ & \ 0.8936(7)$^{\rm b)}$ & \ 0.4659(6)$^{\rm c)}$ & \ 1.3266(11)$^{\rm c)}$ \\
12 & \ 0.1978(5)$^{\rm b)}$ & \ 0.8937(6)$^{\rm b)}$ & \ 0.2072(3)$^{\rm c)}$ & \ 1.3272(11)$^{\rm c)}$ \\
$\infty$ & \ --- & \ 0.8933(6)$^{\rm b)}$ & \ --- & \ 1.3306$^{\rm d)}$ \\
\hline
\end{tabular}
\end{center}
\end{table}

\section{Phase diagram}
\begin{figure}[ht]
\includegraphics[width=0.9\linewidth]{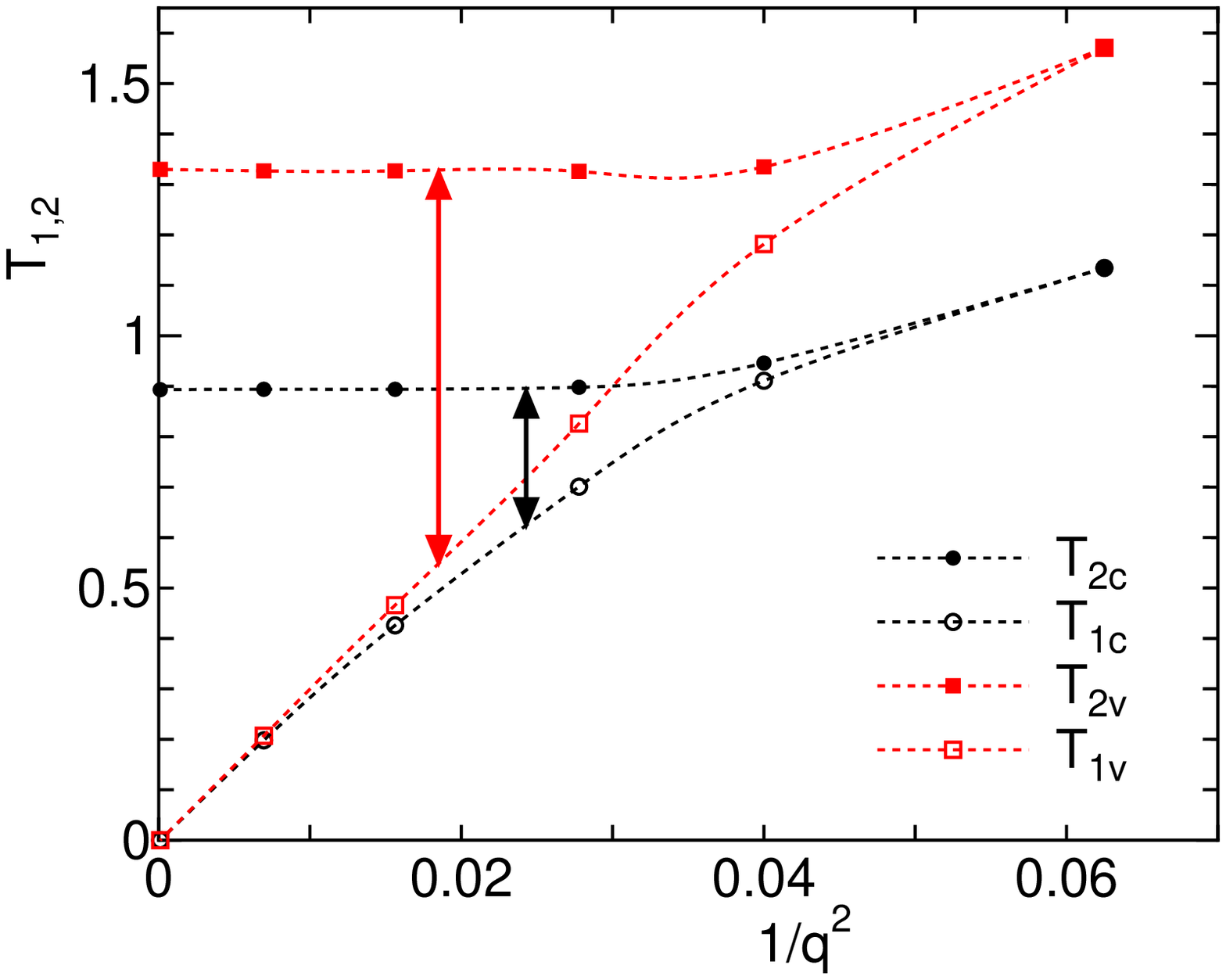}
\caption{(Color online) 
Plot of the $1/q^2$-dependence 
of transition temperatures of regular (cosine) and Villain types 
of clock model. The lower transition temperatures $T_1$ 
and the upper transition 
temperatures $T_2$ are shown by open and filled marks, respectively. 
Dotted curves are guided for the eyes. 
The arrow indicates the BKT phase. 
}
\label{fig_phase}
\end{figure}

We tabulate the critical temperatures of regular and Villain types 
of clock model in Table \ref{TC}. 
In this table we add some estimates of previous references 
\cite{tomita02,takayama,hasenbusch}. 
We plot the $1/q^2$-dependence of $T_1$ and $T_2$ 
of clock models in Fig.~\ref{fig_phase}.  
There, the exact results for $q=4$ are also given; 
for the regular model the Ising singularity occurs 
at $T_c = [\ln(\sqrt{2}+1)]^{-1} = 1.1346$, and  
for the Villain model that occurs at $T_c = \pi/2 = 1.5708$. 
The transition temperature $T_1$ becomes smoothly lower with larger $q$; 
in the lowest order we find that $T_1 \propto 1/q^2$, 
which is consistent with the theoretical prediction \cite{jose}. 
Figure~\ref{fig_phase} is the phase diagram of the clock models. 
The arrow indicates the BKT phase.  We see a narrow BKT phase 
for the regular model of $q=5$.

%--------------------------------fig_phase-----------------------------
%-------------------------------------------------------------------------

\end{document}